\documentstyle[multicol,aps,epsfig,graphicx]{revtex}

\newcommand{\la}{\langle}
\newcommand{\ra}{\rangle}

\newcommand{\mA}{{\mathcal A}}

\newcommand{\var}{{\mathrm Var}}

\begin{document}
\title{Quantum properties of transverse pattern formation in
  second-harmonic generation}

\author{M. Bache$^{1,2,3}$, P. Scotto$^1$, R. Zambrini$^1$,  M. San
  Miguel$^1$ and M. Saffman$^4$}
\address{ 1)
  Instituto Mediterr\'{a}neo de Estudios Avanzados, IMEDEA (SCSIC-UIB), Universitat de les Illes Balears, E-07071 Palma de
  Mallorca, Spain.
  \\
  2) Optics and Fluid Dynamics Department, Ris{\o} National
  Laboratory, Postbox 49, DK-4000 Roskilde, Denmark.
  \\
  3) Informatics and Mathematical Modelling, Technical University of
  Denmark,
  DK-2800 Lyngby, Denmark.\\
  4) Department of Physics,
  University of Wisconsin, 1150 University Avenue,  Madison,
  Wisconsin, 53706, USA.\\
}

\date{\today}
\maketitle
\begin{abstract}
  
  We investigate the spatial quantum noise properties of the one
  dimensional transverse pattern formation instability in intra-cavity
  second-harmonic generation. The Q representation of a
  quasi-probability distribution is implemented in terms of nonlinear
  stochastic Langevin equations.  We study these equations through
  extensive numerical simulations and analytically in the linearized
  limit.  Our study, made below and above the threshold of pattern
  formation, is guided by a microscopic scheme of photon interaction
  underlying pattern formation in second-harmonic generation.  Close
  to the threshold for pattern formation, beams with opposite
  direction of the off-axis critical wave numbers are shown to be
  highly correlated. This is observed for the fundamental field, for
  the second harmonic field and also for the cross-correlation between
  the two fields. Nonlinear correlations involving the homogeneous
  transverse wave number, which are not identified in a linearized
  analysis, are also described.  The intensity differences between
  opposite points of the far fields are shown to exhibit
  sub-Poissonian statistics, revealing the quantum nature of the
  correlations. We observe twin beam correlations in both the
  fundamental and second-harmonic fields, and also nonclassical
  correlations between them.
\end{abstract}

\begin{multicols}{2}

\section{Introduction}
\label{sec:intro}

Pattern formation has been an active area of research in many diverse
systems \cite{cross}. Numerous similarities to pattern formation in
other systems have been reported in recent studies in nonlinear optics
\cite{abraham:1990,lugiato:1994,arecchi:1999,lugiato:1987,etrich:1997}.
However similar, nonlinear optics also displays properties that are
wholly unique due to the relevance of quantum aspects in optical
systems, one manifestation of this is the inevitable quantum
fluctuations of light. In the last decade an effort has been made to
study the interplay in the spatial domain between optical pattern
formation, known from classical nonlinear optics, and the quantum
fluctuations of light \cite{kolobov:1999,lugiato:1999}.  New
nonclassical effects such as quantum entanglement and squeezing in
patterns were predicted \cite{lugiato:1999,lugiato:1997}. Another
interesting example is the phenomenon of quantum images: below the
instability threshold, information about the pattern is encoded in the
way the quantum fluctuations of the fields are spatially correlated
\cite{lugiato:1995}.

Nonlinear $\chi^{(2)}$-materials immersed in a cavity have shown most
promising quantum effects. A paradigm of spatiotemporal quantum
behavior has been the optical parametric oscillator (OPO), which
despite its striking simplicity is able to display highly complex
behavior \cite{lugiato:1993,oppo:1994b,gatti:1997b}.  In the
degenerate OPO, pump photons are down-converted to signal photons at
half the frequency and with a high degree of quantum correlation. This
might be attributed to the fact that the signal photons are created
simultaneously conserving energy and momentum, leading to the notion
of twin photons.  In the opposite process of second-harmonic
generation (SHG) fundamental photons are up-converted to
second-harmonic photons at the double frequency. On a classical level,
both the OPO and intra-cavity SHG display similar spatiotemporal
behavior. The essential difference between them is that in the OPO an
oscillation threshold for the process exists, which simultaneously
acts as the threshold for pattern formation. On the contrary, SHG
always takes place no matter the strength of the pump field, but there
is a threshold that marks the onset of pattern formation. This gives
pronounced differences with the OPO in the linearized behavior below
the threshold for pattern formation. In the OPO the pump and the
signal fields effectively decouple and only the latter becomes
unstable at threshold. At a microscopic level, the behavior of the OPO
close to the threshold can be understood in terms of a unique process
in which a pump photon decays into two signal photons with opposite
wave numbers. In SHG the fundamental and second-harmonic fields are
coupled and both become unstable at threshold. This complicates the
picture mainly by the number of microscopic mechanisms that are
relevant to describe the pattern formation process. But this
complexity, on the other hand, is likely to generate interesting
correlations between the fundamental and the second-harmonic field.
Recently, transverse quantum properties in the singly resonant SHG
setup were investigated \cite{lodahl:2002}.  There, squeezing in the
fundamental output was observed close to the critical wave number, but
since the second-harmonic is not resonated the question of possible
correlations between the two fields was not addressed.  However, since
the second harmonic in the singly resonant case is given directly as a
function of the fundamental, correlations similar to the ones observed
in the fundamental should be expected. In this paper, we will consider
the case of doubly resonant SHG with the aim of investigating the
spatial correlations not only within each field (fundamental field and
second-harmonic field), but also between the two fields.

For this purpose we use the formalism of quasi-probability
distributions \cite{gardiner:2000}. Choosing the use of the Q
representation we are able to derive a set of nonlinear Langevin
equations that describes the time evolution of the quantum fields in
the SHG setup (Sec.~\ref{sec:model}). In Sec.~\ref{sec:linear-appr},
the linear stability analysis of this system will be discussed and a
proper regime of parameters specified, for which the formalism adopted
here is applicable.  Section~\ref{sec:space-corr} will be devoted to
an analysis, on a microscopic level, of the implications of the
three-wave interactions in the nonlinear crystal. These considerations
allow to identify the most important spatial correlations expected in
this two-field system and to define suitable quantities to be
calculated. In particular, we will focus on equal time correlation
functions of intensity fluctuations and we will study photon number
variances when looking for nonclassical features of the intra-cavity
fields. A systematic study of the spatial correlations is presented
first through analytical results in the framework of a linearized
theory below the threshold for pattern formation
(Sec.~\ref{sec:anal}), and also through extensive numerical
simulations of the nonlinear Langevin equations reported below
(Sec.~\ref{sec:corr-below}) and above (Sec.~\ref{sec:corr-above}) the
threshold for pattern formation.  We conclude in
Sec.~\ref{sec:conclusion}.

\begin{figure}
\includegraphics[width=8cm]{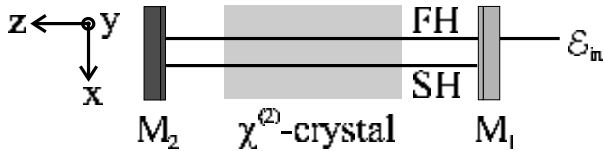}
\caption{The model setup in a top-view.}
\label{fig:model}
\end{figure}

\section{Nonlinear quantum model for intra-cavity SHG}
\label{sec:model}

We consider a nonlinear $\chi^{(2)}$-material with type I phase
matching immersed in a cavity with a high reflection input mirror
$M_1$ and a fully reflecting mirror $M_2$ at the other end, cf.
Fig.~\ref{fig:model}. The cavity is pumped at the frequency $\omega_1$
and through the nonlinear interaction in the crystal photons of
frequency $\omega_2=2\omega_1$ are generated. This is the process of
SHG. The cavity supports a discrete number of longitudinal modes, and
we will consider the case where only two of these modes are relevant,
namely the mode $\omega_{1,{\mathrm cav}}$ closest to the fundamental
frequency and $\omega_{2,{\mathrm cav}}$ closest to the
second-harmonic frequency.  In the setup shown in Fig.~\ref{fig:model}
$\omega_{2,{\mathrm cav}}=2\omega_{1,{\mathrm cav}}$, but we will
allow the cavity resonances to be independent in order to control the
detunings individually. The pump beam propagates along the
$z$-direction and using the mean field approximation, variations in
the $z$-direction are averaged out.  This approach is justified as
long as the losses and detunings are small.  Due to diffraction the
transverse section perpendicular to the $z$-direction spanned by the
$xy$-plane also comes into play. We consider the simple 1D case where
only one of the transverse directions is relevant, so variations along
the $y$-direction are neglected and only the $x$-direction is taken
into account (this could easily be achieved experimentally by using a
crystal with a small height).  Let $\hat A_1(x,t)$ and $\hat A_2(x,t)$
denote the 1D intra-cavity boson operators of the fundamental field
(FH) and second-harminic field (SH), respectively.  They obey the
following equal time commutation relation
\begin{equation}
  \label{eq:Acomm}
  [\hat A_i (x,t),\hat A_j^\dagger(x',t)]=\delta_{ij}\delta(x-x'),
\quad i,j=1,2.
\end{equation}
The Hamiltonian operator describing SHG including diffraction can be
written as done in Ref.~\cite{gatti:1997b} for the OPO,
\begin{equation}
  \label{eq:H}
  \hat H=\hat H_{\mathrm free}+\hat H_{\mathrm int}+\hat H_{\mathrm ext},
\end{equation}
where the free Hamiltonian is given by
\begin{eqnarray}
  \label{eq:Hfree}
    \hat H_{\mathrm free}&=&\hbar \int dx \hat A_1^\dagger(x,t) \left(
  -\delta_1 -
  \frac{c^2}{2\omega_1}\frac{\partial^2}{\partial x^2} \right) \hat
A_1(x,t)
\nonumber \\
&&+ \hbar \int dx \hat A_2^\dagger(x,t) \left( -\delta_2 -
  \frac{c^2}{4\omega_1} \frac{\partial^2}{\partial x^2}\right) \hat A_2(x,t).
\end{eqnarray}
Here $\delta_j=\omega_j-\omega_{j,{\mathrm cav}}$ are the detunings
from the nearest cavity resonances, $\partial^2/\partial x^2$
describes the diffraction, and $c$ is the speed of light. The
interaction Hamiltonian describes the nonlinear interaction in the
material
\begin{equation}
  \label{eq:Hint}
    \hat H_{\mathrm int}=\frac{i\hbar g}{2}\int dx \left(  \hat A_2(x,t)
  ( {\hat A}_1^{\dagger}(x,t))^2 -{\mathrm H.c.}  \right),
\end{equation}
where $g$ is the nonlinear coupling parameter proportional to the
$\chi^{(2)}$-nonlinearity of the crystal. The external Hamiltonian
describes the effects of the pump injected into the cavity at the
fundamental frequency, which is taken to be a classical quantity
${\mathcal E}_{\mathrm in}$, so we have
\begin{equation}
  \label{eq:Hext}
    \hat H_{\mathrm ext}=i\hbar \int dx \left({\mathcal E}_{\mathrm in}
  \hat  A_1^\dagger(x,t) - {\mathcal E}_{\mathrm in}^*  \hat
  A_1(x,t)\right).
\end{equation}
Then the master equation for the density matrix $\hat \rho$ in the
interaction picture is given by
\begin{equation}
  \label{eq:ME}
   \frac{\partial \hat \rho}{\partial
  t}=-\frac{i}{\hbar}[\hat H,\hat \rho]+(\hat L_1+\hat L_2)\hat \rho.
\end{equation}
The cavity losses are assumed to occur only through the input coupling
mirror to the external continuum of modes, and are here included
through the Liouvillian terms
\begin{eqnarray}
  \label{eq:Liouvillian}
  \hat L_j\hat \rho=&&\int dx \gamma_j \Big( 2 \hat A_j(x,t) \hat \rho \hat
  A_j^\dagger(x,t)
\nonumber\\&&
-  \hat \rho \hat A_j^\dagger(x,t) \hat A_j - \hat
  A_j^\dagger(x,t)\hat A_j(x,t)
  \hat\rho \Big),
\end{eqnarray}
where $\gamma_j$ are the cavity loss rates. Here we have assumed that
thermal fluctuations in the system can be neglected.

Using the standard approach of expanding the density matrix into
coherent states weighted by a quasi-probability distribution function,
the master equation (\ref{eq:ME}) is mapped onto a functional
equation, depending on the order for creation and destruction
operators \cite{hillery:1984,schleich:1997}.  For a Hamiltonian that
is quadratic in the field operators this results in a Fokker-Planck
equation, implying that the dynamical evolution of the distribution
function may be modelled by an equivalent set of classical stochastic
Langevin equations. However, due to the contributions of higher order
to the Hamiltonian (\ref{eq:Hint}) problems may arise.  When using the
Wigner representation the evolution equation of the quasi-probability
functional contains third order derivatives, which means that no
equivalent Langevin equations can be found. These third order terms
have been shown to model quantum jump processes \cite{kinsler:1991}.
In some cases these terms can be neglected leading to an approximate
Fokker-Planck form. When using the P or Q representations problems of
negative diffusion in the Fokker-Planck equation come into play
\cite{gardiner:2000}. To avoid negative diffusion in the P
representation, some techniques have been developed where the phase
space is doubled \cite{drummond:1980a,yuen:1986}, but then numerical
problems due to divergent stochastic trajectories generally appear
\cite{gilchrist:1997,zambrini:2001a}.  We choose here to use the Q 
representation which in a restricted domain of parameters has a
nonnegative diffusion matrix and has been shown to be a useful
alternative in the similar problem of calculating nonlinear quantum
correlations in the OPO \cite{zambrini02}. The Q representation has no
singularity problems, is bounded and always nonnegative.

Introducing $\alpha_i$ and $\alpha_i^*$ as the c-number equivalents of
the intra-cavity boson operators $\hat A_i$ and $\hat A_i^\dagger$,
the evolution equation for the quasi-probability distribution function
$Q(\underline\alpha)$ is
\begin{eqnarray}
\label{eq:Q-FPE}
  \frac{\partial Q(\underline \alpha)}{\partial t}&&=
\Bigg(
\frac{\partial}{\partial \alpha_1}[(\gamma_1-i\delta_1)\alpha_1
  -g\alpha_1^*\alpha_2-i\frac{c^2}{2\omega_1}\frac{\partial^2}{\partial x^2}
-{\mathcal E}_{{\mathrm in}}]
\nonumber\\
+&&\frac{\partial}{\partial \alpha_2}[(\gamma_2-i\delta_2)\alpha_2
  +\frac{g}{2}\alpha_1^2-i\frac{c^2}{4\omega_1}\frac{\partial^2}{\partial x^2}]
  -\frac{g}{2}\alpha_2\frac{\partial^2}{\partial\alpha_1^2}
\nonumber\\
+&&\gamma_1\frac{\partial^2}
    {\partial\alpha_1 \partial
    \alpha_1^*} + \gamma_2
    \frac{\partial^2}{\partial\alpha_2
    \partial \alpha_2^*} + {\mathrm c.c.}
 \Bigg)Q(\underline\alpha),
\end{eqnarray}
with $\underline \alpha=\{\alpha_1,\alpha_1^*,\alpha_2,\alpha_2^*\}$.
This is just an extension to the diffractive case of the result
obtained by Savage \cite{savage:1988}.  Equation~(\ref{eq:Q-FPE}) has
the form of a Fokker-Planck equation, and it has positive diffusion if
\begin{eqnarray}\label{pos-cond}
|\alpha_2|< 2 \frac{\gamma_1}{g}.
\end{eqnarray}
As shown below, it is possible to fix the parameters of the system in
such a way that the stable solution for the SH field is well below the
value $2 \gamma_1/g$.  Fluctuations around this stable solution are
small, so that the probability violating the condition
(\ref{pos-cond}) is almost zero. Neglecting then stochastic
trajectories violating this condition, we may write a set of
equivalent Langevin stochastic equations by applying the Ito formalism
for the stochastic integration \cite{ito-stratonovich}. We then obtain the
following nonlinear Langevin equations 
\begin{mathletters}
\begin{eqnarray}
  \label{eq:Lang-unscaled}
    \partial_{t} \alpha_1(x,t)=&&(-\gamma_1
  +i\delta_1)\alpha_1(x,t) +g\alpha_1^*(x,t)\alpha_2 (x,t)
\nonumber\\&&
  +i\frac{c^2}{2\omega_1}\frac{\partial^2}{\partial x^2}\alpha_1(x,t)+{\mathcal E}_{\mathrm in}
+\sqrt{2\gamma_1}\xi_1(x,t), \\
  \partial_t \alpha_2(x,t)=&&(-\gamma_2
  +i\delta_2)\alpha_2(x,t)- \frac{g}{2}\alpha_1^2(x,t)
\nonumber\\&&
 + i\frac{c^2}{4\omega_1}\frac{\partial^2}{\partial x^2}\alpha_2(x,t)
+\sqrt{2\gamma_2}\xi_2(x,t),
\end{eqnarray}
\end{mathletters}
with multiplicative Gaussian white noise sources correlated as
follows
\begin{mathletters}
\begin{eqnarray}
  \label{eq:xiQ_unscaled}
    \la \xi_i^*(x,t)\xi_j(x',t')\ra&=&\delta_{ij}
  \delta(x-x')\delta(t-t'), \\
  \la\xi_2(x,t)\xi_2(x',t')\ra&=&0, \\
  \la \xi_1(x,t)\xi_1(x',t')\ra&=&-\frac{g\alpha_2(x,t)}{2\gamma_1}
  \delta(x-x')\delta(t-t').
\end{eqnarray}
\end{mathletters}
We rescale space and time according to
\begin{equation}
   \tilde t= t\gamma_1, \quad \tilde x= x/l_d,
\label{eq:space_time_norm}
\end{equation}
where $l_d$ is the characteristic length scale given by
\begin{equation}
  \label{eq:ld}
  l_d^2=\frac{c^2}{2\gamma_1\omega_1}.
\end{equation}
We also normalize the fields and noise according to
\begin{eqnarray}
\label{eq:fields_norm}
  &&A_j(x,t)= \alpha_j(x,t)\frac{g}{\gamma_1}, \quad
  \tilde\xi_j(x,t)=\xi_j(x,t)\sqrt{\frac{l_d}{\gamma_1}},
\nonumber\\
  &&E={\mathcal E}_{\mathrm in}\frac{g}{\gamma_1^2}.
\end{eqnarray}
This allows to rewrite the Langevin equations in dimensionless
form:
\begin{mathletters}
\label{eq:Lang-scaled}
\begin{eqnarray}
  \partial_{\tilde t} A_1(\tilde x,\tilde t)&=&(-1 +i\Delta_1)
  A_1(\tilde x,\tilde t)+A_1^*(\tilde x,\tilde t)A_2(\tilde x,\tilde t)
\nonumber\\&&
+i\frac{\partial^2}{\partial \tilde x^2} A_1(\tilde x,\tilde t)+E
+\sqrt{\frac{2}{n_{th}}}\tilde\xi_1(\tilde x,\tilde t), \\
  \partial_{\tilde t} A_2(\tilde x,\tilde t)&=&(-\gamma +i\Delta_2)
  A_2(\tilde x,\tilde t)-\frac{1}{2}A_1^2(\tilde x,\tilde t)
\nonumber\\&&
+\frac{i}{2}\frac{\partial^2}{\partial \tilde x^2} A_2(\tilde x,\tilde t)
  +\sqrt{\frac{2\gamma}{n_{th}}}\tilde\xi_2(\tilde x,\tilde t),
\end{eqnarray}
\end{mathletters}
where $\gamma=\gamma_2/\gamma_1$ and $\Delta_j=\delta_j/\gamma_1$, and
$E$ may be taken real.  Moreover we have introduced
\begin{equation}
  \label{eq:nth}
 n_{th}=\frac{\gamma_1^2l_d}{g^2},
\end{equation}
which in the OPO coincides with the number of photons in the
characteristic ``area'' $l_d$ required to trigger the oscillation.
The noise strength is seen to scale like $n_{th}^{-1/2}$.  The
normalized noise sources are correlated by
\begin{mathletters}
\begin{eqnarray}
  \la \tilde \xi_i^*(\tilde x, \tilde t)\tilde \xi_j(\tilde
  x',\tilde t')\ra&=&\delta_{ij}
  \delta(\tilde x-\tilde x')\delta(\tilde t-\tilde t'), \\
  \la\tilde\xi_2(\tilde x,\tilde t)\tilde\xi_2(\tilde x',\tilde t')\ra&=&0, \\
  \la \tilde\xi_1(\tilde x,\tilde t)\tilde \xi_1(\tilde x',\tilde t')\ra&=&
  -\frac{A_2(\tilde x,\tilde t)}{2}
  \delta(\tilde x-\tilde x')\delta(\tilde t-\tilde t').
\label{eq:noise-mult}
\end{eqnarray}
\end{mathletters}
These noise sources turn out to only to be defined for
\begin{eqnarray}
\label{pos-cond-scale}
|A_2(\tilde x,\tilde t)|<2,
\end{eqnarray}
which coincides with the condition (\ref{pos-cond}) for a positive
diffusion expressed in terms of the rescaled fields.

In the following the tildes are dropped, and only normalized
dimensionless equations are considered. We will also use the
terminology $\omega\equiv \omega_1$ and $2\omega\equiv \omega_2$.

\section{Linearized equations and bifurcation diagram}
\label{sec:linear-appr} 

In this section we consider the linearization of the nonlinear
Langevin equations in the Q representation around the homogeneous
steady state solutions below the threshold for pattern formation. This
approach relies on the assumption that the fluctuations are small with
respect to the field mean values, and therefore we expect this
approach to break down close to the instability threshold.  We will
come back later (Sec.~\ref{sec:num-res}) to the question of the
validity of the linear approximation.  We write the fields as
$A_j(x,t) = {\mathcal A}_j + \beta_j(x,t)$, where $\beta_j(x,t)$
represent the fluctuations around ${\mathcal A}_j$. The classical
homogeneous values ${\mathcal A}_j$ of the fields given by the
homogeneous steady state solutions of the deterministic limit
($n_{th}\rightarrow \infty$) of Eqs.~(\ref{eq:Lang-scaled}), as found
in Ref.\cite{etrich:1997}. Using this in Eqs.~(\ref{eq:Lang-scaled})
we find the following set of linearized equations
\begin{mathletters}
\label{eq:SHG-linear}\\
\begin{eqnarray}
  \partial_t \beta_1(x,t)&=&(-1 +i\Delta_1)
  \beta_1(x,t)+{\mathcal A}_2 \beta_1^*(x,t)
\nonumber\\
  &+& {{\mathcal A}^*_1} \beta_2(x,t)+i  \frac{\partial^2}{\partial x^2}
  \beta_1(x,t)+\sqrt{\frac{2}{n_{th}}}\xi_1(x,t),
\label{eq:SHG-lang-A1-unscaled-Q}\\
  \partial_t  \beta_2(x,t)&=&(-\gamma +i\Delta_2)
  \beta_2(x,t)-{\mathcal A}_1 \beta_1(x,t)
\nonumber\\
&+&\frac{i}{2}
  \frac{\partial^2}{\partial x^2} \beta_2(x,t)+
  \sqrt{\frac{2\gamma}{n_{th}}}\xi_2(x,t).
\label{eq:SHG-lang-A2-unscaled-Q}
\end{eqnarray}
\end{mathletters}
The correlations of the stochastic sources $\xi_i(x,t)$ in the
linearized limit become
\begin{mathletters}
\begin{eqnarray}
  \la \xi_i^*(x,t)\xi_j(x',t')\ra&=&\delta_{ij}
  \delta(x-x')\delta(t-t'),\\
  \la \xi_1(x,t)\xi_1(x',t')\ra&=&-\frac{{\mathcal A}_2}{2}
  \delta(x-x')\delta(t-t'),\label{phasdep:cor}\\
  \la\xi_2(x,t)\xi_2(x',t')\ra&=&0.
  \label{pha:dep:cor:real}
\end{eqnarray}
\end{mathletters}
With ${\mathcal A}_2$ being merely a constant, the noise in the
linear approximation is not multiplicative any more.  However, as
in the nonlinear equations we have the restriction
\begin{equation}
  \label{eq:pos-cond-lin}
|{\mathcal  A}_2|<2.
\end{equation}
We would like to mention that the Wigner representation, in the linear
regime, would lead to equivalent results without suffering from any
limitation since it satisfies a Fokker-Planck equation for any value
of $|{\mathcal A}_2|$.  However, for the sake of a consistent
presentation of our results we have chosen to consider the Q
representaion also in the linear case.

It is instructive to introduce the spatial Fourier transform of the
fluctuations
\begin{equation}
  \beta_j(k,t)=\int_{-\infty}^{\infty} \frac{dx}{\sqrt{2\pi}}
\beta_j(x,t)e^{ik x},
\end{equation}
which physically represents the amplitude of the fluctuations in the
far field.  Considering Eqs.~(\ref{eq:SHG-linear}) and their complex
conjugates, it is readily shown that these amplitudes $\beta_j(k,t)$
fulfill a set of equations which can be written in the following
matrix form
\begin{mathletters}
\label{eq:lang-lin-farfield}
\begin{eqnarray}
&&\partial_t
\left(\begin{array}{c}
  \beta_1(k,t)\\ \beta_1^*(-k,t) \\ \beta_2(k,t) \\
  \beta_2^*(-k,t)
\end{array}\right)
={\mathbf M}(k)
\left(\begin{array}{c}
  \beta_1(k,t)\\ \beta_1^*(-k,t) \\ \beta_2(k,t) \\
  \beta_2^*(-k,t)
\end{array}\right)
\nonumber\\
&&+\sqrt{\frac{2}{n_{th}}}
\left(\begin{array}{c}
  \eta_1(k,t)\\ \eta_1^*(-k,t) \\ \sqrt{\gamma}\eta_2(k,t) \\
  \sqrt{\gamma}\eta_2^*(-k,t)
\end{array} \right),
\label{eq:beta}
\\
&&{\mathbf M}(k)=\left(\begin{array}{cccc}
  \sigma_1(k) & {\mathcal A}_2 & {\mathcal A}_1^{*} & 0 \\
  {\mathcal A}_2^{*} & \sigma_1^*(k) & 0 & {\mathcal A}_1 \\
  -{\mathcal A}_1 & 0 & \sigma_2(k) & 0 \\
  0 & -{\mathcal A}_1^{*} & 0 & \sigma_2^*(k) \\
\end{array}\right),
\label{eq:M}
\end{eqnarray}
\end{mathletters}
where $\sigma_1(k)=-1+i(\Delta_1-k^2)$ and $\sigma_2(k)=-\gamma
+i(\Delta_2-k^2/2)$ have been introduced and each noise term
$\eta_j(k,t)$ is the Fourier transform of the noise term appearing
in the real space linearized Langevin
equations~(\ref{eq:SHG-linear}). Their correlations are given by
\begin{mathletters}
\begin{eqnarray}
  \la \eta_i^*(k,t)\eta_j(k',t')\ra&=&\delta_{ij}
  \delta(k-k')\delta(t-t'),\\
  \la \eta_1(k,t)\eta_1(k',t')\ra&=&-\frac{{\mathcal A}_2}{2}
  \delta(k+k')\delta(t-t'),\\
  \la\eta_2(k,t)\eta_2(k',t')\ra&=&0.
  \label{pha:dep:cor}
\end{eqnarray}
\end{mathletters}

The linear stability of the classical equations obtained as the
$n_{th}\rightarrow \infty$ limit of Eqs.~(\ref{eq:SHG-linear}) was
investigated by Etrich \textit{et al.} \cite{etrich:1997}. A rich
variety of instabilities was shown to exist: A self-pulsing
instability, that leads to oscillations of the homogeneous steady
states without any transverse structure, was present for all
parameters. The oscillatory transverse instability leading to patterns
traveling in space and time was only present for certain parameters
and branched out from the self-pulsing instability.  Bistability was
demonstrated for large detunings of same sign and for $\gamma$ small.
Most importantly, for all parameters also stationary transverse
instabilities were found to exist, \textit{i.e.} instabilities at a
critical transverse wave number $k=k_c$ and with zero imaginary
eigenvalue. It was shown that stripe-type solutions exist but are
always unstable, and numerical simulations showed that instead
hexagons are the dominating stationary transverse instability.  The 1D
configuration we have chosen to consider here has the advantage that
the pattern always will be a stripe and therefore leads to simpler
interpretation of the correlations.  We will choose a range of
parameters in which the stationary transverse instability is
accessible as the primary bifurcation. This bifurcation is
supercritical in the 1D model.

\begin{figure}
\includegraphics[width=8cm]{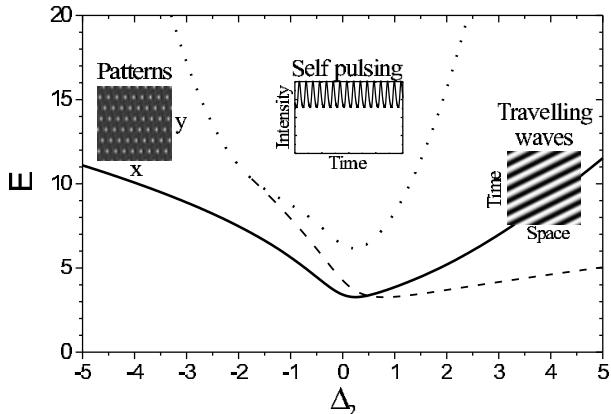}
\caption{Stability diagram for $\Delta_1=2.0$ and $\gamma=0.5$,
showing
  transverse stationary instability (solid line), transverse
  oscillatory instability (dashed line) and self-pulsing instability
  (dotted line).}
\label{fig:Ebif}
\end{figure}

\begin{figure}
\includegraphics[width=8cm]{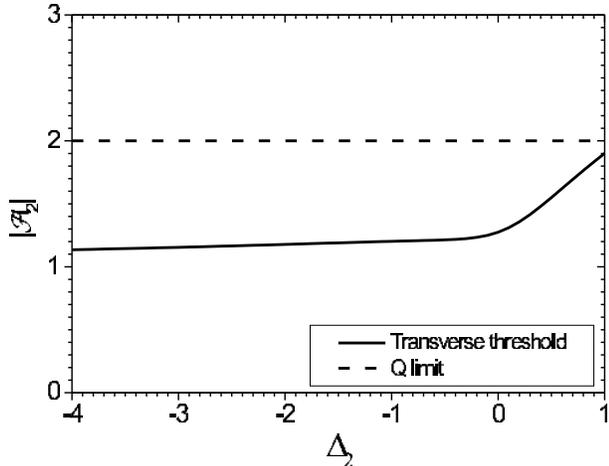}
\caption{Transverse instability for $\Delta_1=2.0$ and
  $\gamma=0.5$ shown for the intra-cavity second-harmonic field,
  along with the limit for the Q representation, $|A_2|<2$.}
\label{fig:qlimit}
\end{figure}

The choice of parameters must take into account the requirement of
applicability of the Q representation.  One finds that
Eq.~(\ref{eq:pos-cond-lin}) can only be satisfied for $\Delta_1>0$
\cite{competing}.  Using the expressions presented in
Ref.~\cite{etrich:1997} and fixing $\Delta_1=2.0$ and $\gamma=0.5$ we
obtain the bifurcation diagram shown in Fig.~\ref{fig:Ebif}
\cite{mathematica}. We observe that for $\Delta_2<0$ it is possible to
obtain stationary patterns (solid line) as the primary bifurcation at
a critical value of the pump, $E_t$; increasing the pump beyond $E_t$
eventually the system will also become self-pulsing unstable (dotted
line). For $\Delta_2>0$ the transverse oscillatory bifurcation (dashed
line) is the primary one, and therefore travelling waves are observed
in this region. The bistable area is located for $\Delta_2>8.3$ and
hence beyond the range shown here.

Expressing the onset of transverse instability, seen in
Fig.~\ref{fig:Ebif}, in terms of the intra-cavity value of the SH we
have the bifurcation diagram for the transverse instability shown in
Fig.~\ref{fig:qlimit}. We see that for $\Delta_2<0$ we are well below
the limit for positive diffusion~(\ref{eq:pos-cond-lin}). Therefore,
the probability of trajectories violating the condition
(\ref{pos-cond-scale}) of the nonlinear equations is almost zero.  For
$\Delta_2>0$, increasing $\gamma$ or decreasing $\Delta_1$ towards
zero, this threshold gets closer to $|{\mathcal A}_2|=2$.

We will therefore use the parameters $\Delta_1=2.0$, $\Delta_2=-2.0$
and $\gamma=0.5$ in the rest of this paper, which gives a pattern
formation threshold of $E_t=7.481757$ and a critical wave number
$k_c=1.833$. The noise strength is set to $n_{th}=10^{8}$ which
is a typical value for the cavity setup discussed here
\cite{unpublished}.

\begin{figure}
\includegraphics[width=8cm]{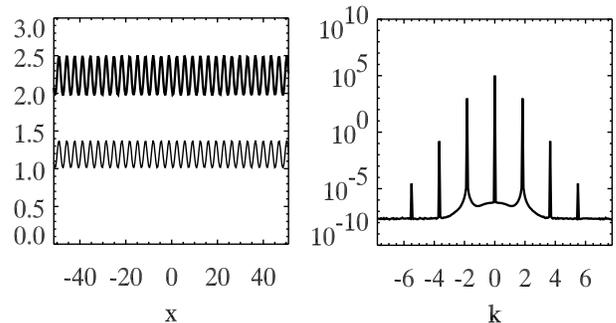}
\caption{Numerical simulation of the Langevin equations above threshold with
  $E/E_t=1.01$ and $L=102.84$.
  Left: The absolute value of the near field of the FH (above) and SH
  (below). Right: Far field average
  intensity of FH, $\la |A_1(k)|^2\ra$. The far field of the SH shows
  a similar structure.}
\label{fig:nearfar}
\end{figure}

The main task of the following section is to identify the most
important correlations we expect to find in the system. For this
purpose it is useful to have a good knowledge of the spatial
structures that emerge in the system. 

Numerical simulations \cite{numerics:quantum} of the nonlinear
Eqs.~(\ref{eq:Lang-scaled}) confirmed the instability at a finite
transverse wave number $k=k_c$ predicted by the linear stability
analysis. Above the threshold for pattern formation modulations was
observed around the steady state with wavelengths corresponding to
$k_c$. This is shown in Fig.~\ref{fig:nearfar} where the far field
intensity shows distinct peaks at $k=0$, corresponding to the
homogeneous background, and at $k=\pm k_c$ corresponding to the
modulations observed in the near field, as well as higher harmonics.

\begin{figure}
  \begin{center}
\includegraphics[width=8.5cm]{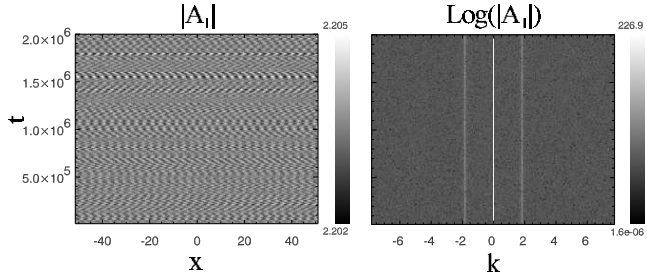}
  \end{center}
\caption{Numerical simulation with $E/E_t=0.9999$ and $L=103.057$,
  showing the space-time evolution of $|A_1|$ in the near field (left)
  and far field (right). A similar behavior is seen for the SH.}
\label{fig:nearfar_below}
\end{figure}

Below threshold the quantum noise will excite the least damped
modes and precursors of the spatial pattern are observed. This is
shown in Fig.~\ref{fig:nearfar_below} where a space-time plot is
presented for the FH near and far field.  Clearly a stripe-type
pattern is formed, but as time progresses the noise diffuses the
pattern \cite{lugiato:1995,zambrini:2000} so that averaging over time
will wash out this emerging structure and a spatially homogeneous
near field will remain. On the contrary, as we will show the
spatial correlation functions do encode precise information about
the emerging pattern, even after this time averaging has been
carried out, as illustrated through the concept of quantum images
\cite{lugiato:1995}.

\section{Correlations, photon interaction and pattern formation}
\label{sec:space-corr}

Our general objective is the investigation of the spatial intra-cavity
field correlations emerging in this system as a result of the coupling
of FH and SH field through the nonlinearity of the crystal, and the
implications of the spatial instability on these correlations.  This
study has a two-fold purpose: First, to obtain a precise picture on
how pattern formation occurs in cavity SHG. In particular, we will aim
at identifying the relevant mechanisms, in terms of elementary
three-wave processes that are important for the understanding of the
intra-cavity field dynamics. Secondly, it will be interesting to
investigate whether these correlations are the manifestation of
nonclassical states of the fields.  Such states are identified by
investigating the statistics of the intra-cavity intensities, looking
in particular for possible sub-Poissonian features
\cite{davidovich:1996}.

\subsection{Photon interaction}
\label{sec:photon-chem}

We will start by investigating the equal time correlations between
intensity fluctuations at different points in the far field. The
intensity of each field being directly proportional to the number of
photons in the corresponding mode, we can relate the intensity
fluctuations to the creation or destruction of photons. The idea is
that the way these fluctuations are correlated gives information about
the microscopic mechanisms that take place in the cavity and,
ultimately, that are involved in the pattern formation process.
Generally speaking, a positive correlation tells us that there should
exist a coherent mechanism that creates simultaneously the
corresponding photons. The following normalized correlations are
considered
\begin{equation}
C_{ij}^n(k,k')= \frac{\la \delta
\hat N_i(k,t) \delta \hat N_j(k',t)\ra}{\sqrt{\la \delta
\hat N_i(k,t)^2\ra\la \delta \hat N_j(k',t)^2\ra}},
\label{eq:corr-int}
\end{equation}
where the superscript $n$ denotes normalization. The intensity
fluctuations are given by $\delta \hat N_j(k,t)=\hat N_j(k,t)-\la \hat
N_j(k,t)\ra$, which involves the photon number operator $\hat
N_j(k,t)=\hat A_j^\dagger (k,t) \hat A_j(k,t)$.  The brackets denote
quantum mechanical averages (expectations) of the operators, which in
our approach based on numerical simulations of equivalent c-numbers
will be translated into an average over time. The normalization of the
correlations implies that $C_{ij}^n(k,k')=1$ for perfectly correlated
fluctuations, whereas $C_{ij}^n(k,k')=-1$ will be the signature of
perfect anti-correlation between the intensity fluctuations. As usual,
the absence of any correlation will translate into a vanishing
correlation function $C_{ij}^n(k,k')=0$.  In the following we will
refer to $C^n_{11}(k,k')$ and $C^n_{22}(k,k')$ as self-correlations
(between different modes of a given field) and to $C^n_{12}(k,k')$ as
cross-correlations (between modes in different fields).

\begin{figure}[t]
  \begin{center}
 \includegraphics[width=8cm]{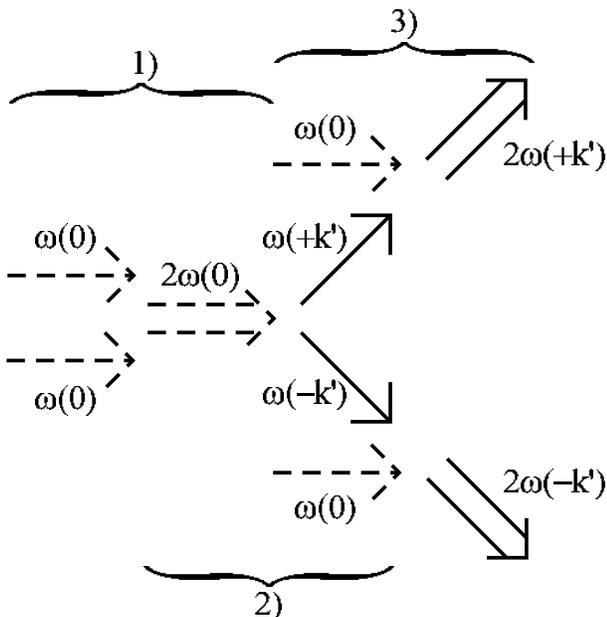}
 \caption{The basic picture of pattern formation on a microscopic
   level through SHG. The single arrows ($\rightarrow$) symbolize FH
   photons, while double arrows ($\Rightarrow$) symbolize SH photons.
   The dashed arrows are photons from the homogeneous background.}
 \label{fig:basics}
\end{center}
\end{figure}

As a guideline for the investigation of the properties of these
correlation functions, the first step consists in identifying the
basic photon processes when the system is taken close to a transverse
instability. These photon processes must obey the standard energy and
momentum conservation laws. Whereas the former merely implies that
each elementary process must connect one SH photon with two FH
photons, the latter will translate into a condition on the transverse
wave numbers. Keeping in mind that the cavity is pumped with an
homogeneous field at the frequency $\omega$, the first process to
consider consists in two homogeneous FH photons, $[\omega](k=0)\equiv
[\omega](0)$ combining to give one homogeneous SH photon,
$[2\omega](0)$, what will be written as
$[\omega](0)+[\omega](0)\rightarrow [2\omega](0)$.  This is encoded in
the Hamiltonian term $\hat A_1^2\hat A_2^\dagger$ in
Eq.~(\ref{eq:Hint}).  The inverse process, which corresponds to the
degenerate OPO process, also takes place in the system, as shown by
the presence of the term $(\hat A_1^\dagger)^2\hat A_2$ in
Eq.~(\ref{eq:Hint}).  Elaborating on these considerations we propose
the scheme in Fig.~\ref{fig:basics} as the simplest way of obtaining a
pattern in both fields.
\begin{itemize}
\item [1)] The first step is the basic SHG channel where two
  homogeneous FH photons give a SH photon and vice versa,
  \textit{i.e.} the channel $[\omega](0)+[\omega](0)\leftrightarrow
  [2\omega](0)$. It is important to realize that fluctuations around
  the steady state are considered, hence it is not considered how the
  FH photons combine to give the steady state SH photons via the
  channel above, but rather how the fluctuations invoke the channel
  beyond this.
\item [2)] The second step is the down-conversion of a SH photon into
  two FH photons. Momentum conservation in the process implies that
  the two FH photons have the same value of the transverse wave number
  but with opposite signs. These are called twin photons since an
  emission of a $[\omega](+k')$ photon must be accompanied by an
  emission of a $[\omega](-k')$ photon, and they therefore show a high
  degree of correlation. This channel written as
  $[2\omega](0)\leftrightarrow[\omega](-k')+[\omega](+k')$ generates
  off-axis FH photons.
\item [3)] Off-axis SH photons are obtained by combining the created
  off-axis FH photon from step 2) with a photon from the homogeneous
  background to give a SH photon, which by momentum conservation must
  have the same wave number as the off-axis FH photon. This channel
  can be written as $[\omega](0)+[\omega](+k')\leftrightarrow
  [2\omega](+k')$.
\end{itemize}

Of course, these are not the only three-wave processes which are
kinematically allowed in the nonlinear crystal, since the interaction
Hamiltonian (\ref{eq:Hint}) induces any process of the form
$[\omega](k')+[\omega](k'')\leftrightarrow [2\omega](k'+k'')$, with
arbitrary wave numbers $k'$ and $k''$. In fact, the basic scheme we
propose in Fig.~\ref{fig:basics} only takes into account those
three-wave processes which involve at least one photon of the
homogeneous background fields.  Empirically, this choice is motivated
by the observation that below the threshold these are the only field
modes that are macroscopically populated, so that any process
involving them should be stimulated in analogy to what occurs in
standard stimulated emission.  Formally, the selection of these
particular elementary processes corresponds precisely to the
approximation made by linearizing the field equations around the
steady state solution. As can easily be checked, the full equations
for the far field fluctuations contain additional terms quadratic in
the fluctuation amplitudes, which indeed account for other three-wave
processes.  Linearizing we are left with Eq.~(\ref{eq:beta}), which
only take into account the processes represented by step 2) and step
3).  These processes translate into nondiagonal elements of the matrix
${\bf M}(k)$ of the linear system, and as a consequence, for any value
of $k$, the time evolution of the four amplitudes $\beta_1(k,t)$,
$\beta_1(-k,t)$, $\beta_2(k,t)$ and $\beta_2(-k,t)$ will be coupled.
This coupling is expected to translate into correlations between the
intensity fluctuations $\delta I_1(k)$, $\delta I_1(-k)$, $\delta
I_2(k)$, $\delta I_2(-k)$.

This preliminary observation already allows to give a more explicit
interpretation of the basic scheme of Fig.~\ref{fig:basics}.
Splitting the dynamics of the intra-cavity fields into independent
elementary steps, as suggested in the discussion of
Fig.~\ref{fig:basics}, would not explain any correlations either
between $[\omega](k')$ and $[2\omega](-k')$ nor between
$[2\omega](k')$ and $[2\omega](-k')$.  Hence the inspection of the
linearized equations shows that the interpretation of
Fig.~\ref{fig:basics} in terms of a cascade is too naive. Instead, we
have to understand step 2) and 3) as two coherent, joint processes,
which generate simultaneously correlations between the 4 modes
$[\omega] (k')$, $[\omega] (-k')$, $[2\omega](k')$ and $[2\omega]
(-k')$.  Finally, it is important to stress that the linearized
analysis does not predict any correlation between intensity
fluctuations in field modes with wave numbers of different modulus.
Mathematically, this is due to the fact that in the linear
approximation all correlation functions (\ref{eq:corr-int}) have the
structure
\begin{equation}
C_{ij}(k,k')=C_{ij}^{(-)}(k)\delta^2(k-k')
+C_{ij}^{(+)}(k)\delta^2(k+k'),
\label{eq:corr-int2}
\end{equation}
as will be shown in the next section.  Close enough to threshold,
however, this will not be true any more because of the emergence of
additional correlations of nonlinear nature.

Let us finally briefly address the fundamental difference between OPO
and SHG: Whereas in SHG, the two fields ${\mathcal A}_1$ and
${\mathcal A}_2$ are always nonzero regardless of the pump level, in
the OPO case below the oscillation threshold ${\mathcal A}_2$ is fixed
by the pump and ${\mathcal A}_1=0$. Considering the scheme presented
in Fig.~\ref{fig:basics}, the vanishing of ${\mathcal A}_1$ implies
that there is no macroscopic population of the mode $[\omega] (0)$ and
therefore, step 3) of Fig.~\ref{fig:basics} is not present. The route
to pattern formation simply consists of step 2) in
Fig.~\ref{fig:basics}, generating correlations between $\delta \hat
N_1(k,t)$ and $\delta \hat N_1(-k,t)$.  Mathematically, the
consequence for the stability of the homogeneous solution is that the
two equations (\ref{eq:SHG-linear}) effectively decouple and that only
the FH becomes unstable at the threshold.

\subsection{Correlations below shot-noise}
\label{sec:SN-corr}

Once correlations between intensity fluctuations are identified, it is
interesting to investigate if they are connected to nonclassical
states of the intra-cavity fields. A coherent field obeys Poissonian
photon statistics, which implies that the variance and the mean of the
photon number operator $\hat N$ are equal. Let us consider the photon
number operators associated with the sum and difference of the
intensities at different far-field points $\hat N_i(k)\pm \hat
N_j(k')$, where $\hat N_i(k)=\hat a_i^\dagger(k) \hat a_i(k)$ and
$\hat N_j(k)=\hat a_j^\dagger(k) \hat a_j(k)$ are the number operators
of two states $\hat a_i(k,t)$ and $\hat a_j(k,t)$. Since we will
consider equal time correlation functions in the steady state of the
system, from now on we will drop the time argument of the field
operators. Taking out the special case $i=j$ and $k'=k$ which will be
treated separately, the variance expressed in normal order (indicated
by dots) reads
\begin{eqnarray}
  {\mathrm Var}(\hat N_i(k)&\pm \hat N_j(k'))=:\!{\mathrm Var}(\hat
  N_i(k)\pm \hat N_j(k'))\!:
\nonumber\\
&+\la:\! \hat N_i(k)\!:\ra[\hat a_i(k),\hat a_i^\dagger(k)]\nonumber\\
& +\la:\!
  \hat N_j(k')\!:\ra[\hat a_j(k'),\hat a_j^\dagger(k')],
  \label{eq:var-P}
\end{eqnarray}
where ${\mathrm Var}(X)\equiv \la X^2\ra-\la X\ra^2$. For a coherent
state, the normal ordered variance vanishes, and the mean, given by
last two terms in (\ref{eq:var-P}), represents the shot noise level
for the considered quantity
\begin{equation}
  \label{eq:SN}
  S.N.=\la :\!\hat N_i(k)\!:\ra[\hat a_i(k),\hat a_i^\dagger(k)] +\la
  :\! \hat N_j(k')\!:\ra [\hat a_j(k'),\hat a_j^\dagger(k')].
\end{equation}
If the normal ordered variance becomes negative
\begin{equation}
  \label{eq:squeezing-crit}
  :\!{\mathrm Var}(\hat N_i\pm \hat N_j)\!:<0,
\end{equation}
the variance becomes less than the mean, indicating sub-Poissonian
behavior. Such a nonclassical state is identified when the
correlation normalized to the shot-noise level, defined as
\begin{eqnarray}
 \lefteqn{C_{ij}^{(\pm)}(k,k')\equiv}\hspace{8cm}\nonumber\\
\frac{:\!{\mathrm Var}
[\hat N_i(k)\pm \hat N_j(k')]\!:}{\la :\!\hat N_i(k)\!:\ra
[\hat a_i(k),\hat a_i^\dagger(k)] +\la :\!  \hat
  N_j(k')\!:\ra [\hat a_j(k'),\hat a_j^\dagger(k')] }+1,
  \label{def:cpm}
\end{eqnarray}
is such that $C_{ij}^{(\pm)}(k,k')<1$.  The computation of this
quantity requires to write the normal ordered quantities appearing in
(\ref{def:cpm}) in terms of anti-normal ordered quantities, since
these are the quantities that are computed as averages in our Langevin
equations associated with the Q representation. Using the identities
\begin{mathletters}
\begin{eqnarray}
  \vdots \hat N_i(k)\vdots&=&\hat a_i(k)\hat a^\dagger_i(k)=:\!\hat N_i(k)\!:
+[\hat a_i(k),\hat a^\dagger_i(k)],
\label{eq:N_anti}\\
  \vdots \hat N^2_i(k)\vdots&=&\hat a_i(k)\hat a_i(k)\hat a^\dagger_i(k) 
\hat a^\dagger_i(k)
\nonumber\\&=&:
\!\hat N_i(k)^2\!:+4:\!\hat N_i(k)\!:[\hat a_i(k),\hat a^\dagger_i(k)]
\nonumber\\&&+
2[\hat a_i(k),\hat a^\dagger_i(k)]^2
\end{eqnarray}
\end{mathletters}
with three dots indicating anti-normal ordering, Eq.~(\ref{def:cpm})
reads, when expressed in terms of anti-normal ordered quantities
\end{multicols}
\begin{equation}
  \label{eq:var-Q-ab_SN}
  C_{ij}^{(\pm)}(k,k')
=\frac{\vdots{\mathrm Var}[\hat N_i(k)\pm
  \hat N_j(k')]\vdots-\la\vdots
  \hat N_i(k)\vdots\ra[\hat a_i(k),\hat a^\dagger_i(k)]-\la\vdots \hat N_j(k') 
\vdots\ra
  [\hat a_j(k'),\hat a^\dagger_j(k')]}{ \la \vdots\hat N_i(k)\vdots 
\ra[\hat a_i(k),\hat a^\dagger_i(k)] +
\la \vdots\hat N_j(k') \vdots \ra[\hat a_j(k'),\hat a^\dagger_j(k')]
  -[\hat a_i(k),\hat a^\dagger_i(k)]^2-[\hat a_j(k'),\hat a^\dagger_j(k')]^2}.
\end{equation}
\begin{multicols}{2}
Then \textit{e.g.} the normalized correlation ${\mathrm Var}[\hat N_1(k)\pm
\hat N_1(-k)]/{S.N.}$ may be found by setting $i=j=1$ and $k'=-k$. 
Equation~(\ref{eq:var-Q-ab_SN}) is valid for $k,k'\neq 0$, while the 
special case $k=0$ will be addressed in the specific cases.

\section{Linearized calculations below threshold}
\label{sec:anal}

Below threshold, the linear approximation scheme allows to derive
semi-analytical expressions for the correlation functions defined in
the previous section. These may be expressed in terms of the auxiliary
correlation function
\begin{eqnarray}
C_{ij}^Q(k,k')&=&{\langle \vdots\delta \hat N_i(k,t)\delta \hat
  N_j(k',t)\vdots\rangle},
\quad
    i,j=1,2\nonumber\\
 &=&{\langle | A_i(k,t)|^2 | A_j(k',t)|^2\rangle}\nonumber\\
&&-{\langle | A_i(k,t)|^2 \rangle\langle| A_j(k',t)|^2\rangle},
\label{rappel:def}
\end{eqnarray}
where the superscript Q indicates that the average is done with the
Q representation, corresponding to anti-normal ordered quantities, as
indicated in the first line of (\ref{rappel:def}).

The starting point of our analysis is the set of linearized Langevin
equations (\ref{eq:beta}) which have the exact solutions
\begin{eqnarray}
&&\left(\begin{array}{c}
  \beta_1(k,t)\\ \beta_1^*(-k,t) \\ \beta_2(k,t) \\
  \beta_2^*(-k,t)
\end{array}\right)
=
e^{{\mathbf M}(k)t}
\left(\begin{array}{c}
  \beta_1(k,0)\\ \beta_1^*(-k,0) \\ \beta_2(k,0) \\
  \beta_2^*(-k,0)
\end{array}\right)
\nonumber\\
&&+ \sqrt{\frac{2}{n_{th}}}e^{{\mathbf M}(k)t}\int_0^t dt'
e^{-{\mathbf M}(k)t'}
\left(\begin{array}{c}
  \eta_1(k,t')\\
  \eta_1^*(-k,t')\\
  \sqrt{\gamma}\eta_2(k,t')\\
  \sqrt{\gamma}\eta_2^*(-k,t')
\end{array}\right).
\label{eq:formal}
\end{eqnarray}
The first term in Eq.~(\ref{eq:formal}) describes how the intra-cavity
fields with arbitrary initial conditions relax to the steady state
solution and it does not contribute to the steady state correlations.
The second term in Eq.~(\ref{eq:formal}) gives the response of the
intra-cavity fields to the vacuum fluctuations entering the cavity
through the partially transparent input mirror.  Starting from
Eq.~(\ref{eq:formal}), it is possible to derive semi-analytical
expressions for the correlations (\ref{rappel:def})
\begin{eqnarray}
C^Q_{ij}(k,k')
&=&
\langle | \beta_i(k,t)|^2 | \beta_j(k',t)|^2
\rangle
\nonumber\\
&-&
\langle | \beta_i(k,t)|^2\rangle\langle
| \beta_j(k',t)|^2
\rangle
\nonumber\\
&+&2 {\mathrm Re} \big\{ {\mathcal A}_i^* {\mathcal A}_j \langle
\beta_i(k,t)\beta^*_j(k',t)\rangle
\nonumber\\
&&
+{\mathcal A}_i^*{\mathcal A}_j^* \langle \beta_i(k,t)
\beta_j(k',t)\rangle
\big\}
\delta(k)\delta(k'),
\label{generexpr}
\end{eqnarray}
where ${\mathrm Re}\{\cdot\}$ denotes the real part.  Whereas the
first two terms in the right-hand side (r.h.s.) of
Eq.~(\ref{generexpr}) measure the correlations in the intensities of
the fluctuations, the last two terms can be traced back to
interferences between the fluctuations and the homogeneous component
of each field.  Since these interferences only contribute to the equal
time correlations when $k = k' =0$, we will first concentrate on $k,k'
\neq 0$ and come back later to this special case. Henceforth, unless
otherwise specified we consider the case $k,k'\neq 0$.

The Gaussian character of the fluctuations in this linearized Langevin
model allows to factorize (\ref{generexpr}) in terms of second order
moments of the field fluctuations
\begin{eqnarray}
C^Q_{ij}(k,k')=| \langle \beta_i(k,t) \beta^*_j(k',t)\rangle|^2+
| \langle \beta_i(k,t) \beta_j(k',t)\rangle|^2.
\label{eq:factorize}
\end{eqnarray}
The field correlations $\langle \beta_i(k,t) \beta^*_j(k',t)\rangle$
and $\langle \beta_i(k,t) \beta_j(k',t)\rangle $ can be best evaluated
for the solution Eq.~(\ref{eq:formal}) if we introduce the set of
eigenvectors $\{{\mathbf v}^{(l)}(k)\}_{l=1,...,4}$ of the matrix
${\mathbf M}(k)$, defined through:
\begin{equation}
{\mathbf M}(k) {\mathbf v}^{(l)}(k)=\lambda^{(l)}(k) {\mathbf v}^{(l)}(k).
\end{equation}
An arbitrary $4$-component vector ${\mathbf w}$ can be decomposed on
this basis
\begin{equation}
{\mathbf w}(k)=\left(\begin{array}{c}
  w_1(k)\\w_2(k)\\w_3(k)\\w_4(k) \end{array}\right)
=\sum_{l=1}^{4}w^{(l)}(k){\mathbf v}^{(l)}(k),
\end{equation}
and its components $w^{(l)}$ in the new basis are calculated via the
linear transformation
\begin{equation}
w^{(l)}(k)=\sum_{m=1}^{4}T_{lm}(k)w_m(k).
\end{equation}
This involves a $4\times 4$ matrix $T_{lm}(k)$ calculated as
${\mathbf T}(k)={\mathbf V}(k)^{-1}$ with $V_{lm}(k)=v^{(m)}_{l} (k)$.
Decomposing now the noise vector appearing on the r.h.s. of
Eq.~(\ref{eq:formal}) on this basis
\begin{equation}
\left(\begin{array}{c}
  \eta_1(k,t')\\
  \eta_1^*(-k,t')\\
  \sqrt{\gamma}\eta_2(k,t')\\
  \sqrt{\gamma}\eta_2^*(-k,t')
\end{array}\right)=\sum_{l=1}^{4}\eta^{(l)}(k,t'){\mathbf v}^{(l)}(k),
\end{equation}
allows to rewrite Eq.~(\ref{eq:formal}) in the large time
limit as
\begin{eqnarray}
\left(\begin{array}{c}
  \beta_1(k,t)\\ \beta_1^*(-k,t) \\ \beta_2(k,t) \\
  \beta_2^*(-k,t)
\end{array}\right)
&&= \sqrt{\frac{2}{n_{th}}}\int_0^t dt'
\nonumber\\
&&\times\sum_{l=1}^{4}
e^{{\lambda}^{(l)}(k)(t-t')}\eta^{(l)}(k,t'){\mathbf v}^{(l)}(k).
\label{eq:dec}
\end{eqnarray}
The needed field correlations are given as
\end{multicols}
\begin{mathletters}
\begin{eqnarray}
\langle \beta_i(k,t) \beta_j^*(k',t) \rangle &=&
\frac{2}{n_{th}}\int_0^t dt'
\int_0^t dt''\sum_{l,m=1}^{4}v_{2i-1}^{(l)}(k){v_{2j-1}^{(m)}}^*(k')
e^{{\lambda}^{(l)}(k)(t-t')}e^{{\lambda^{(m)}}^*(k')(t-t'')}
\langle \eta^{(l)}(k,t')
{\eta^{(m)}}^*(k',t'')\rangle, \label{example1}\\
\langle \beta_i(k,t) \beta_j(k',t) \rangle &=&
\frac{2}{n_{th}}\int_0^t dt'
\int_0^t dt''\sum_{l,m=1}^{4}v_{2i-1}^{(l)}(k)v_{2j-1}^{(m)}(k')
e^{{\lambda}^{(l)}(k)(t-t')}e^{{\lambda}^{(m)}(k')(t-t'')}
\langle \eta^{(l)}(k,t')
\eta^{(m)}(k',t'')\rangle.
\label{example2}
\end{eqnarray}
\label{eq:example}
\end{mathletters}
\begin{multicols}{2}
The noise correlations in the new basis $\langle \eta^{(l)}(k,t')
{\eta^{(m)}}^*(k',t'')\rangle$ and $\langle \eta^{(l)}(k,t')
\eta^{(m)}(k',t'')\rangle$ are
\begin{mathletters}
\begin{eqnarray}
\langle \eta^{(l)}(k,t')
{\eta^{(m)}}^*(k',t'')\rangle&=&A_{lm}(k)\delta(k-k')
\delta(t'-t''),\label{corr:new1}\\
\langle \eta^{(l)}(k,t')
\eta^{(m)}(k',t'')\rangle&=&B_{lm}(k)\delta(k+k')
\delta(t'-t'').\label{corr:new2}
\end{eqnarray}
\label{eq:corr-new}
\end{mathletters}
where the matrix elements of the $4\times 4$ matrices ${\mathbf A}(k)$
and ${\mathbf B}(k)$ can easily be evaluated in terms of the matrix
elements $T_{lm}\equiv T_{lm}(k)$ as
\begin{mathletters}
\begin{eqnarray}
  A_{lm}(k)=&& T_{l1}T^{*}_{m1}-\frac{{\mathcal A}_2}{2}
T_{l1}T^*_{m2}+T_{l2}T^{*}_{m2}
\nonumber\\
&&-\frac{{\mathcal A}^*_2}{2}
T_{l2}T^*_{m1}+ \gamma T_{l3}T^{*}_{m3}+\gamma T_{l4}T^{*}_{m4},
  \label{eq:A}
\\
  B_{lm}(k)=&& T_{l1}T_{m2}-\frac{{\mathcal A}_2}{2}
T_{l1}T_{m1}+T_{l2}T_{m1}
\nonumber\\
&&-\frac{{\mathcal A}^*_2}{2} T_{l2}T_{m2}
+\gamma T_{l3}T_{m4}+\gamma T_{l4}T_{m3}.
\label{eq:B}
\end{eqnarray}
\end{mathletters}
Inserting Eqs.~(\ref{eq:corr-new}) in Eqs.~(\ref{eq:example}) we can
easily carry out the time integration, and neglecting transient
contributions, we end up with the following expressions
\begin{mathletters}
\begin{eqnarray}
\lim_{t\to \infty} \langle \beta_i(k,t)\beta^*_j(k',t)
\rangle&=& \frac{2}{n_{th}}
G_{ij}^{(-)} (k)\delta(k-k'),\\
\lim_{t\to \infty} \langle \beta_i(k,t )\beta_j(k',t)\rangle&=&
\frac{2}{n_{th}}
G_{ij}^{(+)} (k)\delta(k+k').
\end{eqnarray}
\end{mathletters}
with
\begin{mathletters}
\label{eq:gpm}
\begin{eqnarray}
G_{ij}^{(-)}(k)=\sum_{l=1}^4\sum_{m=1}^4
A_{lm}(k)\frac{v_{2i-1}^{(l)}(k){v_{2j-1}^{(m)}}^*(k)}
{-(\lambda^{(l)}(k)+{\lambda^{(m)}}^*(k))},\label{eqgpl}\\
G_{ij}^{(+)}(k)=\sum_{l=1}^4\sum_{m=1}^4
B_{lm}(k)\frac{v_{2i-1}^{(l)}(k)v_{2j-1}^{(m)}(k)}
{-(\lambda^{(l)}(k)+\lambda^{(m)}(k))}\label{eqgm}.
\end{eqnarray}
\end{mathletters}
In terms of $G_{ij}^{(-)}(k)$ and $G_{ij}^{(+)}(k)$,
Eq.~(\ref{eq:factorize}) is given by
\begin{eqnarray}
C^Q_{ij}(k,k')=&& \frac{4}{n_{th}^2}\big[| G_{ij}^{(-)}(k)|^2
\delta^2(k-k')
\nonumber\\
&&+| G_{ij}^{(+)}(k)|^2 \delta^2(k+k')\big].
\label{corr:un}
\end{eqnarray}

\subsection{Intensity fluctuation correlations}
\label{sec:ff-fluc-corr} 
It is now easy to compute the normalized correlation function
Eq.~(\ref{eq:corr-int}).  This involves taking into account the
commutation relation Eq.~(\ref{eq:Acomm}) which reads 
\begin{eqnarray}
\label{eq:Comm-scaled} \big[{\hat A_i}(k, t),{\hat A_j}^\dagger
(k',t)\big]= \delta_{ij}\frac{1}{n_{th}}\delta(k-k'),
\end{eqnarray}
after rescaling space and time according to
Eq.~(\ref{eq:space_time_norm}) and the operators similar to the
c-number fields in Eq.~(\ref{eq:fields_norm}). We finally find
\begin{eqnarray}
C^{n}_{ij}(k,k')=&& \frac{| G_{ij}^{(-)}(k)|^2}
{\sqrt{\eta_i(k)}\sqrt{\eta_j(k)}}
\frac{\delta(k-k')^2}{\delta(0)^2}
\nonumber\\
&&+
\frac{| G_{ij}^{(+)}(k)|^2} {\sqrt{\eta_i(k)}\sqrt{\eta_j(k)}}
\frac{\delta(k+k')^2}{\delta(0)^2},
\label{exp:norm:corr}
\end{eqnarray}
with $\eta_j(k)=G_{jj}^{(-)}(k)(G_{jj}^{(-)}(k)-1/2)$, the $-1/2$ in
the parenthesis reflecting the conversion from anti-normal to direct
ordering.  Unlike the mathematical expression (\ref{exp:norm:corr})
derived for an ideally infinite system, the correlation functions
determined from the simulations will have peaks of a finite width,
which will be determined by the discretization in $k$-space used in
the numerical codes, \textit{i.e.} the inverse of the total length of the
system.  This difference, however, will not alter the only relevant
information, which is the height of each of these peaks.  In fact, the
quantities
\begin{mathletters}
\label{exp:norm:strength}
\begin{eqnarray}
C^{n}_{jj}(k,-k)&=&\frac{| G_{jj}^{(+)}(k)|^2}
{{\eta_j(k)}}\\
C^{n}_{12}(k,\pm k)&=&\frac{| G_{12}^{(\mp)}(k)|^2}
{\sqrt{\eta_1(k)}\sqrt{\eta_2(k)}},
\end{eqnarray}
\end{mathletters}
characterize the strength of the correlations between the modes
$[\omega](k)$ and $[\omega](-k)$, $[2\omega](k)$ and $[2\omega](-k)$,
$[\omega](k)$ and $[2\omega](k)$, and $[\omega](k)$ and
$[2\omega](-k)$ respectively.  One easily checks that
$C^{n}_{ii}(k,k)=1$, as a result of an autocorrelation.

All the expressions derived so far are only valid for nonvanishing
transverse wave numbers. At $k=k'=0$, we already observed that there
are extra contributions to the equal time correlation function, as
expressed by Eq.~(\ref{generexpr}).  Furthermore, in the framework of
an expansion in the small parameter $\sqrt{2/n_{th}}$, it is obvious
that these extra terms even dominate, since they scale with
$|\beta_i(k,t)|^2 \sim 2/n_{th}$, whereas the contributions on the
first line of Eq.~(\ref{generexpr}) scales with $|\beta_i(k,t)|^4\sim
(2/n_{th})^2$.  Hence, in the leading order, the correlation function
at $k=k'=0$ is given by
\begin{eqnarray}
C_{12}^Q(k,k')&&\big|_{k=k'=0}
=\frac{2\delta(0)}{n_{th}} 2 {\mathrm Re} \Big( {\mathcal A}_1^*
{\mathcal A}_2 G_{12}^{(-)}(0)
\nonumber\\
&&+{\mathcal A}_1^*{\mathcal A}_2^* G_{12}^{(+)}(0)\Big)
\delta(k)\delta(k')\big|_{k=k'=0}.
\end{eqnarray}
Similar calculations as before allow us to derive the following
expression for the value of the normalized cross-correlation at
$k=k'=0$
\begin{equation}
C^{n}_{12}(0,0)=\frac{{\mathrm Re} \left( {\mathcal A}_1^*{\mathcal A}_2
    G_{12}^{(-)}(0)+{\mathcal A}_1^* {\mathcal A}_2^*G_{12}^{(+)}(0)
    \right)}{\sqrt{ \zeta_1}\sqrt{\zeta_2}},
\label{eq:CN_kzero}
\end{equation}
where $\zeta_j=| {\mathcal A}_j|^2 (G_{jj}^{(-)}(0)-1/4)
+{\mathrm Re}\{{{\mathcal A}_j^*}^{2} G_{jj}^{(+)}(0)\}$.

\subsection{Nonclassical photon number variances}
\label{sec:sn-ana} 

The photon number variances considered in Sec.~\ref{sec:SN-corr} can
be calculated in terms of the auxiliary functions $G_{ij}^{(-)}(k)$
and $G_{ij}^{(+)}(k)$ as well.  The anti-normal ordered quantities in
Eq.~(\ref{eq:var-Q-ab_SN}) can be directly calculated by averages in
the Langevin equation, so below threshold the anti-normal ordered
variance is for $k\neq 0$
\begin{equation}
\vdots\var[\hat N_i(k)\pm \hat N_j(-k)]\vdots =
\var[| \beta_i(k,t)|^2 \pm | \beta_j(-k,t)|^2 ].
\label{eq:varAN}
\end{equation}
Using the commutation relations~(\ref{eq:Comm-scaled}), the
commutators in Eq.~(\ref{eq:var-Q-ab_SN}) are
$[a_j(k),a^\dagger_j(k)]=\delta(0)/n_{th}$, and the normalized
self-correlations takes the form
\begin{equation}
C_{jj}^{(\pm)}(k,-k)=
\frac{2\left(| G_{jj}^{(-)}(k)|^2 \pm
| G_{jj}^{(+)}(k)|^2\right)-
G_{jj}^{(-)}(k)} {G_{jj}^{(-)}(k)-1/2}.
\end{equation}
Similarly, the cross-correlations are
\end{multicols}
\begin{equation}
C_{12}^{(\pm)}(k,\nu k)=
\frac{2\left(\sum_j| G_{jj}^{(-)}(k)|^2 \pm 2| G_{12}^{(-\nu)}(k)|^2\right)-
\sum_j G_{jj}^{(-)}(k)}
{\sum_j G_{jj}^{(-)}(k)-1}, \quad \nu=+1,-1
\end{equation}

When $k=k'=0$ Eq.~(\ref{eq:varAN}) is no longer valid. Instead,
following the procedure outlined for the normalized correlations
we have to the leading order $O(n_{th}^{-1})$
\begin{equation}
  \label{eq:C_SN_kzero}
  C_{12}^{(\pm)}(0,0)=4\frac{{\mathrm Re}\left[
\sum_{j}{\mA_j^*}^2G_{jj}^{(+)}(0)\pm 2
\mA_1^*\left(\mA_2^*G_{12}^{(+)}(0) +\mA_2G_{12}^{(-)}(0)\right)
\right] + \sum_{j}|\mA_j|^2 G_{jj}^{(-)}(0)}{\sum_j|\mA_j|^2}-1.
\end{equation}
\begin{multicols}{2}
The self-correlations become
\begin{mathletters}
  \label{eq:SC_sn_kzero}
\begin{eqnarray}
  C_{jj}^{(-)}(0,0)&=&0,\label{eq:CSN_self_minus}\\
  C_{jj}^{(+)}(0,0)&=&4{\mathrm Re}\left[
  e^{-i2\phi_{A_j}}G_{jj}^{(+)}(0) \right] + 4G_{jj}^{(-)}(0) -1.
\end{eqnarray}
\end{mathletters}
where $\phi_{A_j}$ is the phase of ${\mathcal A}_j$. Note that
$C_{jj}^{(+)}(0,0)$ is actually $\var[\hat N_j(0)]$ normalized to
shot-noise. The result of Eq.~(\ref{eq:CSN_self_minus}) is simply
because the correlation $C_{jj}^{(-)}(k,k')$ amounts to calculating
the variance of zero for $k=k'=0$.

\begin{figure}[t]
  \begin{center}
 \includegraphics[width=8.5cm]{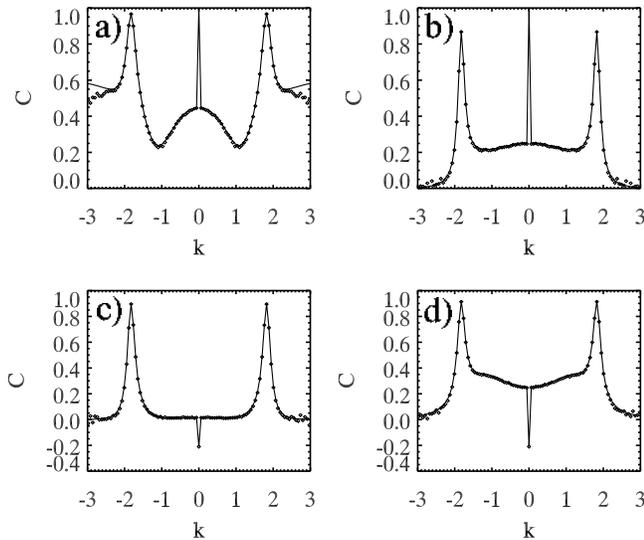}
 \caption{The linear self-correlations a) $C^n_{11}(k,-k)$, b)
   $C^n_{22}(k,-k)$ and linear cross-correlations c) $C^n_{12}(k,k)$, d)
   $C^n_{12}(k,-k)$ as function of the transverse wave number for
   $E/E_t=0.99$. The points are numerical results while the lines are
   analytical results.}
 \label{fig:C_num_an}
\end{center}
\end{figure}

\section{Correlations below threshold}
\label{sec:corr-below}

The linearized results of Sec.~\ref{sec:anal} gives an analytical
insight to the behaviour below threshold for pattern formation.
However, very close to the threshold this linear approximation breaks
down because of critical nonlinear fluctuations, and additional
contributions may emerge as for example shown in a vector Kerr model
by Hoyuelos \textit{et al.}  \cite{hoyuelos:1998}. Such nonlinear
correlations can be calculated through numerical simulations of the
full nonlinear evolution equations. 

In this section we present numerical results obtained from simulations
of the nonlinear equations~(\ref{eq:Lang-scaled}) below threshold,
with the parameters discussed in Sec.~\ref{sec:linear-appr}.  Our
numerical results are compared with the analytical results of the
previous section, and therefore also serves as a cross-check of our
analytical and numerical methods.

\subsection{Linear correlations: Analytical and numerical results}
\label{sec:anal-corr-theory}

We first consider the strength of the correlations between symmetric
points in the far fields below the threshold for pattern formation. In
Fig.~\ref{fig:C_num_an}, the four quantities defined by
Eq.~(\ref{eq:corr-int}) are plotted. The data are obtained from
numerical simulations and from the analytical results of
Eqs.~(\ref{exp:norm:strength}) and~(\ref{eq:CN_kzero}). Very good
agreement is found between numerics and analytical results.

There are three main features to be considered in the results of
Fig.~\ref{fig:C_num_an}. First, all curves present a distinctly peaked
behavior around the critical wave number $k_c$ for pattern formation,
which means that the corresponding modes are more strongly correlated
than the modes at any other wave number.  Manifestly, this behavior is
connected with the pattern formation mechanism and is closely related
to the phenomenon of quantum images \cite{lugiato:1995}. Secondly, we
also note that in all four plots the correlations show a jump at
$k=0$.  In a) and b) it is the trivial manifestation of an
auto-correlation, since for $k=0$, $k$ and $-k$ coincide, while in c)
and d) the jump is due to the extra interferences with the homogeneous
background fields as predicted from Eq.~(\ref{eq:CN_kzero}). Finally,
we observe that the peaks localized around $k_c$ are
superimposed onto smooth correlation profiles.

The strong correlations appearing between the modes associated with
wave numbers around $k_c$ indicate a strongly synchronized emission of
photons in the modes $[\omega] (+k)$, $[\omega](-k)$ and $[2\omega]
(+k)$, $[2\omega](-k)$.  This behavior reflects the direction of
instability of the system.  As a matter-of-fact, regardless that all
transverse modes of both fields are equally excited by the vacuum
fluctuations entering the cavity, the fluctuations of the intra-cavity
field modes around the critical wave vector will be less damped than
the fluctuations in the other modes.  The closer to the threshold, the
more the behavior of the intra-cavity fields will be dominated by the
mode that becomes unstable at the threshold and gives rise to the
pattern.  In the 4-dimensional phase space spanned by the fluctuation
amplitudes
$\{\beta_1(k,t),\beta_1^*(-k,t),\beta_2(k,t),\beta_2^*(-k,t)\}$, this
mode is characterized by a vector with a given direction.  What we
learn from the correlation functions is that the emerging instability
results in an almost perfectly synchronized emission of photons in the
modes $[\omega] (+k)$, $[\omega](-k)$ and $[2\omega] (+k)$,
$[2\omega](-k)$.

\begin{figure}[t]
  \begin{center}
 \includegraphics[width=8cm]{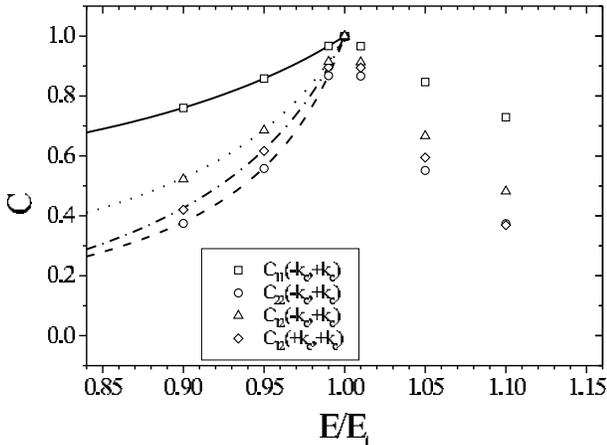}
 \caption{The self-correlations $C^n_{11}(k_c,-k_c)$ (full,
   squares) and $C^n_{22}(k_c,-k_c)$ (dashed, circles) and the
   cross-correlations $C^n_{12}(k_c,-k_c)$ (dotted, triangles) and
   $C^n_{12}(k_c,k_c)$ (dash-dotted, diamonds) as function of the pump
   normalized to the threshold. The points are numerical results while
   the lines are analytical results.}
 \label{fig:CLbelow}
\end{center}
\end{figure}

The dominance of this particular mode when the threshold is approached
is confirmed by the study of the strength of these correlations as a
function of the pump.  In Fig.~\ref{fig:CLbelow} we follow the height
of the peaks at $k=k_c$ of the four linear correlations displayed in
Fig.~\ref{fig:C_num_an}, as a function of the pump level $E/E_t$.  The
most immediate observation is that all the correlations become perfect
in the limit $E\rightarrow E_t$.  This asymptotic behavior can be
understood from the linearized fluctuation analysis presented in
Sec.~\ref{sec:anal}.  It is enough to observe that Eqs.~(\ref{eq:gpm})
involve the inverse of the real part of the eigenvalues of the linear
system (\ref{eq:beta}).  The dominance at the threshold of the
undamped eigenmode of the linear system (\ref{eq:beta}) emerges from
the fact that here the real part of the associated eigenvalue
precisely goes to zero.  Thus, the decrease in the correlations as we
move away from threshold can be seen as the result of the coexistence
of different eigenmodes. Physically the emergence of these
correlations is much less intuitive than the ones in an OPO.  As a
matter-of-fact, in the OPO below the threshold momentum conservation
is enough to predict the existence of correlations between the
fluctuations in the modes $[\omega] (+k)$ and $[\omega](-k)$.  In the
presence of the 4-mode interaction of SHG, the momentum conservation
gives a global condition involving all four beams (at $[\omega] (+k)$,
$[\omega](-k)$ and $[2\omega] (+k)$, $[2\omega](-k)$). These
correlations in fact arise in connection with the emergence of an
instability.

\begin{figure}[t]
  \begin{center}
 \includegraphics[width=8cm]{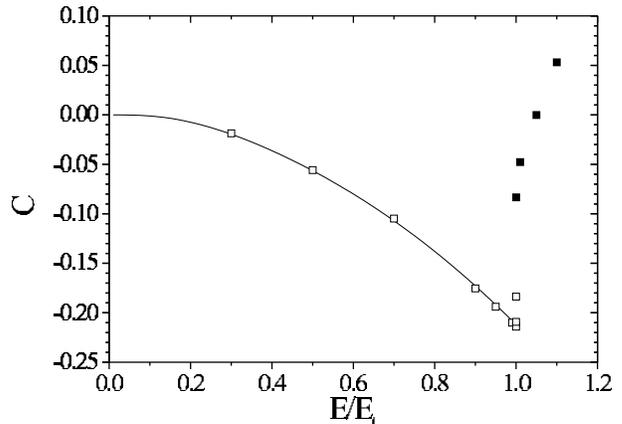}
 \caption{The linear cross-correlation $C^n_{12}(k=0,k'=0)$ as function
   of the pump normalized to the threshold, comparing numerical
   results (points) with the analytical result (line). Open (full) symbols
   are numerics below (above) $E_t$.}
 \label{fig:CXL-hom-below}
\end{center}
\end{figure}

Turning now to the cross-correlation between the homogeneous
components of the fields, we observe that $C^n_{12}(k=0,k'=0)$ in
Fig.~\ref{fig:C_num_an} is negative, reflecting an anticorrelation of
the photons associated with the FH and SH homogeneous waves. In other
words, the creation of a photon $[2\omega](0)$ implies the destruction
of (two) photons $[\omega](0)$ and vice versa.  The origin of this
correlation is much simpler to understand than the previous one: The
two modes $[\omega](0)$ and $[2\omega](0)$ being macroscopically
populated, the vacuum fluctuations simply induce transitions between
these two modes, according to step 1) in the scheme in
Fig.~\ref{fig:basics}. In Fig.~\ref{fig:CXL-hom-below} we plot this
correlation as a function of the pump. Comparing the value of the
correlations below and above threshold, we observe that very close to,
but below, the threshold, the tendency of the curve is reversed and it
anticipates the behavior of the correlation above threshold. These are
nonlinear correlation effects that will be discussed in
Sec.~\ref{sec:num-res}.

Finally, we would like to discuss the smooth contributions to the
correlations displayed in Fig.~\ref{fig:C_num_an}. We first note that
these are not connected with the pattern instability. This was checked
by considering very low pump values for which the peaks around $k_c$
completely vanish, while the smooth structures of the curves remain.
Considering the central region of the curves, roughly for $|k|<k_c$,
the most striking observation is the absence of correlations between
the fluctuations in the modes $[\omega](k)$ and $[2\omega](k)$,
whereas $[\omega](k)$ and $[2\omega](-k)$ are correlated, as well as
$[2\omega](k)$ with $[2\omega](-k)$.  This behavior seems to indicate
the existence of a symmetry restoring principle in the dynamics of the
intra-cavity fields.  As a matter-of-fact, the absence of correlations
between $[\omega](k)$ and $[2\omega](k)$ implies that the fluctuations
of the numbers of pair productions through step 2) and the
fluctuations of the number of conversions $[\omega](k) \to
[2\omega](k)$ through step 3) occur independently of each other.
However, while step 2) of Fig.~\ref{fig:basics} conserves the
$k\rightarrow -k$ symmetry of the system, step 3) does not. As a
consequence, a positive fluctuation in the number of times step 3)
occurs ($[\omega](k)+[\omega](0)\rightarrow [2\omega](k)$),
automatically implies that there will be less $[\omega](k)$ than
$[\omega](-k)$ in the system, and more $[2\omega](k)$ than
$[2\omega](-k)$. The correlations observed may indicate that the
system will try to restore the $k\rightarrow -k$ symmetry by
down-converting $[2\omega](0)\rightarrow [\omega](k)+[\omega](-k)$,
producing a surplus of $[\omega](-k)$ which again will produce more
$[2\omega](-k)$. These mechanisms seems to fit well with the relative
strengths of the correlations observed in the central region of
Fig.~\ref{fig:C_num_an}.  The strongest is always $C^n_{11}(k,-k)$, in
agreement with the fact that the twin photon emission is the principal
source of correlations in the system. Weaker is the correlation
$C_{12}^n(k,- k)$ and even weaker $C_{22}^n(k,-k)$. This
interpretation is consistent with the way the correlations at $k=k_c$
depart from the value $1$ at threshold, when the pump is lowered, as
displayed in Fig.~\ref{fig:CLbelow}.

\begin{figure}[t]
  \begin{center}
 \includegraphics[width=8cm]{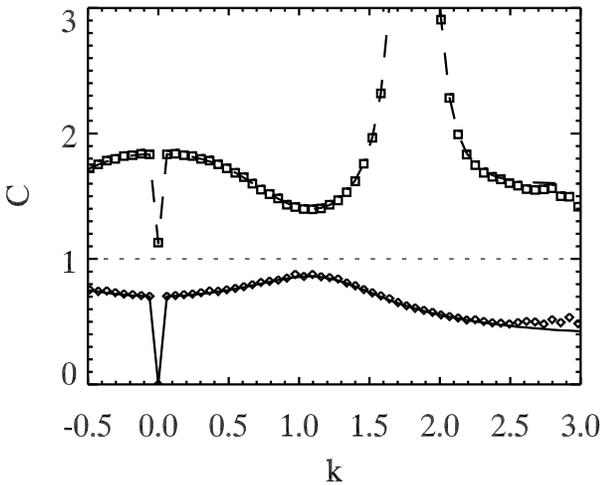}
 \caption{Photon number variances for $E/E_t=0.99$ showing
   $C_{11}^{(-)}(k,-k)$ (full, diamonds) and $C_{11}^{(+)}(k,-k)$
   (dashed, squares). The lines are analytical results while the
   points are numerical simulations. The shot-noise level $C=1$ is indicated
   by a thin dotted line.}
 \label{fig:CSN_FH_below}
\end{center}
\end{figure}

We now turn our attention to the study of the fluctuations in the sum
and difference of the photon numbers at symmetrical points of the far
field. We first consider the twin beam photon variances for the FH,
$C_{11}^{(\pm)}(k,-k)$ defined in Eq.~(\ref{def:cpm}) and shown in
Fig.~\ref{fig:CSN_FH_below}. The results are symmetric with respect to
the substitution $k\rightarrow -k$, wherefore we plotted this quantity
for positive $k$, shifting the origin for better view of the specific
behavior at $k=0$. The linearized calculation predicts sub shot-noise
statistics in the difference $\hat N_1(k)-\hat N_1(-k)$ for all wave
numbers.  For large wave numbers the analytical result for the
correlation approaches the value $1/2$. It is interesting to keep in
mind that for the OPO, the same quantity is equal to $1/2$
independently of the wave number \cite{graham:1984,lane:1988}. In the
SHG case additional processes taking place in the cavity result in a
smooth $k$-dependence of $C_{11}^{(-)}(k,-k)$.  These characteristics
do not depend much on the value of pump, and are not changed
significantly even when the pump level is taken beyond threshold, cf.
Sec.~\ref{sec:corr-above}.  Therefore, the statistics of the intensity
difference is not directly affected by the pattern formation
mechanism. A radically different situation occurs for the
sum-correlation $C_{11}^{(+)}(k,-k)$, which shows a strong peak around
$k=k_c$. For the pump value used in Fig.~\ref{fig:CSN_FH_below} the
peaks correspond to a maximum value $C_{11}^{(+)}(k_c,-k_c)\simeq 35$.
This behavior is connected with the increase of the fluctuations in
the modes associated with the pattern instability when the threshold
is approached, leading to a large excess noise in the statistics of
the intensity of the individual modes $[\omega](k)$ and
$[\omega](-k)$.  This excess noise in each intensity cancels when the
difference $\hat N_1(k)-\hat N_1(-k)$ is considered leading to a
sub-Poissonian statistics, while it is still present in the sum $\hat
N_1(k)+\hat N_1(-k)$. For large $k$ the correlation approaches 1.5,
coinciding again with the corresponding value for the OPO.  Finally,
as before, the jumps at $k=0$ are due to contributions from the
homogeneous steady states, cf.
Eqs.~(\ref{eq:C_SN_kzero})-(\ref{eq:SC_sn_kzero}).

\begin{figure}[t]
  \begin{center}
 \includegraphics[width=8cm]{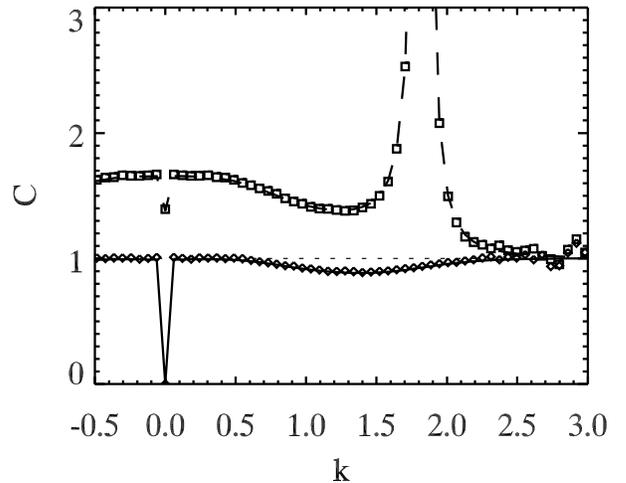}
 \caption{Photon number variances for $E/E_t=0.99$ showing
   $C_{22}^{(-)}(k,-k)$ (full, diamonds) and $C_{22}^{(+)}(k,-k)$
   (dashed, squares).}
 \label{fig:CSN_SH_below}
\end{center}
\end{figure}

The corresponding photon number variances $C_{22}^{(\pm)}(k,-k)$ for
the SH field are shown in Fig.~\ref{fig:CSN_SH_below}. In contrast to
the FH correlations there is almost no sub-shot noise behavior in the
difference correlation $C_{22}^{(-)}(k,-k)$. In other words, the SH
beams only display very weak nonclassical correlations. As for the FH
field, the emerging instability does not influence the noise level in
$C_{22}^{(-)}(k,-k)$, but $C_{22}^{(+)}(k,-k)$ displays a large amount
of excess noise in the vicinity of $k_c$.  The asymptotic large $k$
behavior for both correlations $C_{22}^{(-)}(k,-k)$ and
$C_{22}^{(+)}(k,-k)$ is analytically found to correspond to the
shot-noise limit 1.0.

\begin{figure}[t]
  \begin{center}
 \includegraphics[width=8cm]{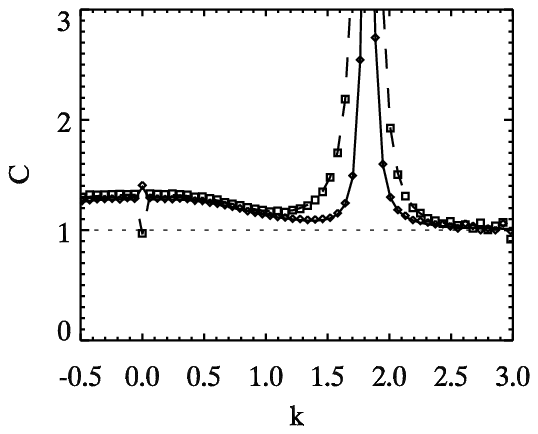}
 \caption{Photon number variances for $E/E_t=0.99$ showing
   $C_{12}^{(-)}(k,k)$ (full, diamonds) and $C_{12}^{(+)}(k,k)$
   (dashed, squares).}
 \label{fig:CSN_FHSH_below}
\end{center}
\end{figure}

The cross-correlations $C_{12}^{(-)}(k,k)$ and $C_{12}^{(+)}(k,k)$ are
shown in Fig.~\ref{fig:CSN_FHSH_below}.  The linearization approach
predicts that these correlations are always above the shot-noise
limit.  Furthermore, at small wave numbers we note that the variances
of the sum and difference coincide. This can only occur when the
fluctuations in the individual modes $[\omega](k)$ and $[2\omega](k)$
are uncorrelated, what was indeed observed in Fig.~\ref{fig:C_num_an}.
Moreover, both the sum and difference correlations show a large excess
noise at $k=k_c$, which is slightly weaker for the difference, as the
result of a partial noise cancellation.

\begin{figure}[t]
  \begin{center}
 \includegraphics[width=8cm]{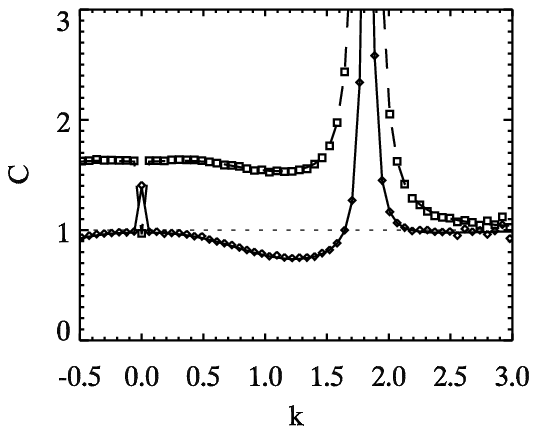}
 \caption{Photon number variances for $E/E_t=0.99$ showing
   $C_{12}^{(-)}(k,-k)$ (full, diamonds) and $C_{12}^{(+)}(k,-k)$
   (dashed, squares).}
 \label{fig:CSN_FHSH_kmk_below}
\end{center}
\end{figure}

The cross-correlations $C_{12}^{(-)}(k,-k)$ and $C_{12}^{(+)}(k,-k)$
are shown in Fig.~\ref{fig:CSN_FHSH_kmk_below}, and here the
difference correlations interestingly go below the shot-noise limit as
long as $k$ is not too close to the critical wave number. It is worth
pointing out that the difference $\hat N_1(k)-\hat N_2(-k)$ shows
nonclassical behavior while the difference $\hat N_1(k)-\hat N_2(k)$
(shown in Fig.~\ref{fig:CSN_FHSH_below}) does not. This somehow paradoxical
situation is related to what was observed in the normalized
correlations where the cross-correlation between $\hat N_1(k)$ and
$\hat N_2(-k)$ was stronger than the almost vanishing
cross-correlation between $\hat N_1(k)$ and $\hat N_2(k)$.  At $k=k_c$
a large amount of excess noise dominates the behavior of both the sum
and the difference correlation and the two correlations show a
pronounced peak.  For large $k$ the correlations approach the shot
noise limit, as seen for the other cross-correlations in
Fig.~\ref{fig:CSN_FHSH_below}.

Olsen \textit{et al.} \cite{olsen:1999} have investigated the
system without spatial coupling corresponding to our results at
$k=0$, and they find that, for certain detunings, the variance of
the sum of the FH and SH intensities are more strongly quantum
correlated than the variance of the individual intensities, due to
the anti-correlation between them. $\var[\hat N_1(0)]/S.N.$ and
$\var[\hat N_2(0)]/S.N.$ can be seen from
Fig.~\ref{fig:CSN_FH_below} and~\ref{fig:CSN_SH_below},
respectively, at $k=0$. Both are larger than the $\var[\hat
N_1(0)+\hat N_2(0)]/S.N.$ observed in
Figs.~\ref{fig:CSN_FHSH_below} and \ref{fig:CSN_FHSH_kmk_below},
so that our results confirm the ones of \cite{olsen:1999}.

\subsection{Nonlinear correlations: Numerical results}
\label{sec:num-res}

So far we have only considered the correlations predicted by the
linearized equations. In order to go beyond this regime, we use our
numerical simulations to search for nonlinear fingerprints in the
correlations and in particular for the emergence of new correlations,
\textit{i.e.} $C^n_{ij}(k,k')$ with $k\neq\pm k'$.  Of particular
interest is to look for correlations between the homogeneous steady
states ($k=0$) and the states with $k=\pm k_c$, $C^n_{ij}(0,\pm k_c)$.
From a technical point of view this task turned out to be difficult
because nonlinear contributions to the correlation functions were only
observable for pump values extremely close to threshold, in a region
where the characteristic time of the dynamics diverges because of
critical slowing down.  This translates into very long transients and
the need of equally long simulations.

\begin{figure}[t]
  \begin{center}
 \includegraphics[width=8cm]{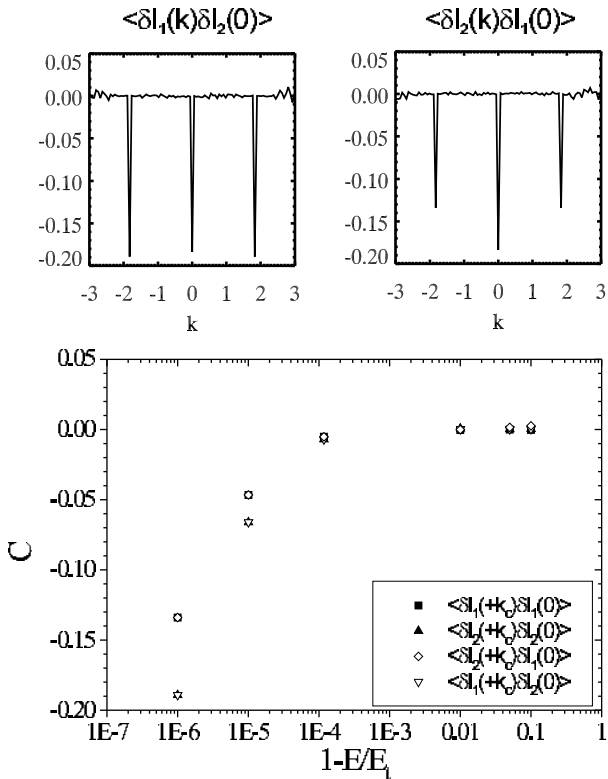}
 \caption{Above: Nonlinear cross-correlations $C^n_{12}(k,k'=0)$ (left) and
   $C^n_{12}(k'=0,k)$ (right) as function of $k$ for $E/E_t=0.999999$.
   Below: Semi-log plot the nonlinear correlations
   $C^n_{ij}(k=+k_c,k'=0)$ as function of $E/E_t$.}
 \label{fig:CNLbelow}
\end{center}
\end{figure}

We have observed some indication of nonlinear correlations for a pump
$E/E_t=0.99999$, which became very clear when using $E/E_t=0.999999$.
For this value of the pump, we show in Fig.~\ref{fig:CNLbelow} our
results for $C^n_{12}(k,k'=0)$ and $C^n_{12}(k'=0,k)$: these curves
put into evidence an anti-correlation between the modes $[\omega](\pm
k_c)$ and $[2\omega](0)$, and between $[2\omega](\pm k_c)$ and
$[\omega](0)$ .  They present a very sharp peak structure, with a
width determined by the distance between two adjacent points of the
discretized $k$-space used for the simulations.  This is due to the
fact that we now consider the correlation functions at fixed $k'$ and
let $k$ vary.  These correlations are a result of nonlinear
amplification of the diverging fluctuations as the threshold is
approached. The negative nature of the correlation is connected with
the fact that the fields with nonzero average values (here the
homogeneous components) act as a "reservoir" of photons for all
processes occurring in the cavity. As we will show later, they are a
precursor of the behavior of the correlations above the threshold. The
correlations at $k=0$ correspond to the linear correlation shown in
Fig.~\ref{fig:CXL-hom-below}. The bottom plot in
Fig.~\ref{fig:CNLbelow} shows the nonlinear correlations
$C^n_{ij}(k=+k_c,k'=0)$ as the threshold is approached.  The
correlations are nonzero only for $E/E_t>0.9999$. As the nonlinear
correlations set in the nonlinear channels in step 2) and 3) of
Fig.~\ref{fig:basics} become stronger and this weakens the
correlations induced by the channel of step 1), which is exactly what
we observed in Fig.~\ref{fig:CXL-hom-below}; $C^n_{12}(0,0)$ becomes
less correlated very close to the threshold. Moreover, we see that the
correlations $C^n_{11}(0,+k_c)$ and $C^n_{12}(0,+k_c)$ have almost
identical values, and the same holds for $C^n_{22}(+k_c,0)$ and
$C^n_{12}(+k_c,0)$. This interesting behavior can be traced back to
the fact that close to the threshold the fluctuations $\delta I_1(k_c)$
and $\delta I_2(k_c)$ are perfectly correlated, as displayed by
Fig.~\ref{fig:CLbelow}, whereas the slight anti-correlation between
$\delta I_1(0)$ and $\delta I_2(0)$ is responsible for the lower
values of $C^n_{22}(+k_c,0)$ and $C^n_{12}(+k_c,0)$ with respect to
$C^n_{11}(0,+k_c)$ and $C^n_{12}(0,+k_c)$.

\section{Correlations above threshold}
\label{sec:corr-above}

Above the threshold for pattern formation the linearized equations
(\ref{eq:SHG-linear}) are no longer valid.  As displayed in
Fig.~\ref{fig:nearfar}, above the threshold not only the homogeneous
modes, but also all modes with wave numbers $k= \pm k_c,\pm 2 k_c, \pm
3 k_c, \ldots$, will present a macroscopic photon number.  Linearizing
around the steady state pattern solution above the threshold under the
assumption of small fluctuations, one obtains new linear equations for
the far field fluctuation amplitudes, which take into account
three-wave processes such as
$[2\omega](k_c)\leftrightarrow[\omega](k)+[\omega](k_c-k)$ or
$[2\omega](k)\leftrightarrow[\omega](k_c)+[\omega](k-k_c)$. In analogy
to the situation below the threshold a linear fluctuation analysis
above the threshold predicts, in addition to the correlations already
present below the threshold, the existence of additional correlations
between the fluctuations $\delta I_1(k)$ and $\delta I_1(k_c-k)$, and
between $\delta I_2(k)$ and $\delta I_1(k-k_c)$.  We will not report
here the explicit results of this cumbersome linear analysis and
restore directly to the numerical analysis of the full nonlinear
Langevin equations.
\begin{figure}[t]
  \begin{center}
 \includegraphics[width=8.5cm]{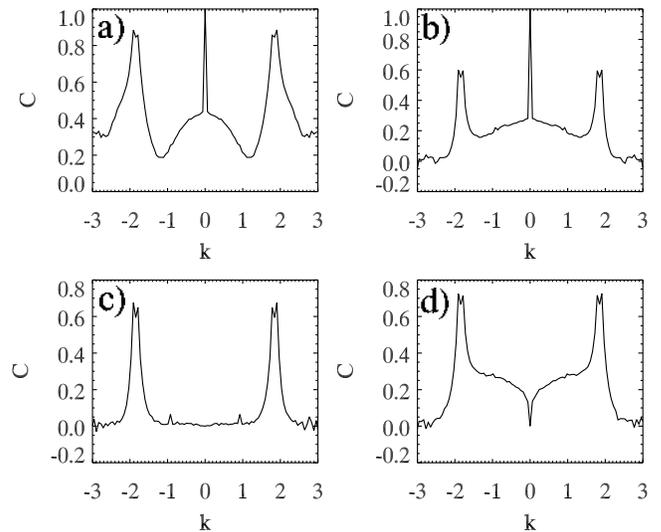}
 \caption{The self-correlations a) $C^n_{11}(k,-k)$, b)
   $C^n_{22}(k,-k)$ and cross-correlations c) $C^n_{12}(k,k)$, d)
   $C^n_{12}(k,-k)$ as function of the transverse wave number for
   $E/E_t=1.05$.}
 \label{fig:C_num_above}
\end{center}
\end{figure}

To investigate the implications of the new field configuration above
the threshold on the intensity correlations, we first consider the
correlations $C^n_{ij}(k,k')$. The same normalized correlations
discussed in Fig.~\ref{fig:C_num_an} below the threshold are plotted
in Fig.~\ref{fig:C_num_above} for a pump value above the threshold. We
observe that the correlations at $k=\pm k_c$ decrease from their
threshold value and are no longer perfect as they were at the
threshold. A closer look actually reveals a dip in the correlations
exactly at the pixels corresponding to $k=\pm k_c$. A tentative
explanation for this is based on the fact that now the modes at the
critical wave number have a finite average value, connected with
macroscopic photon numbers in these modes, whereas the neighbouring
pixels are significantly less populated, cf. the far field of
Fig.~\ref{fig:nearfar}. In comparison the normalized correlations
$C^n_{ij}(k,k')$ show a much smoother behaviour around $k_c$.
Hence, the observed reductions in the correlations above threshold
at $k=\pm k_c$ are connected with spontaneous population exchanges
between these macroscopically populated modes.

In Fig.~\ref{fig:CLbelow} the peaks at $k=\pm k_c$ of
Fig.~\ref{fig:C_num_above} are followed as function of the pump.  The
behavior is very similar to what is seen below the threshold.  Close
to the threshold the correlations are perfect, and as the pump is
taken further away from $E_t$ the correlations become weaker. Below
the threshold this was explained through an eigenvalue competition,
while above the threshold the explanation is that the competitions
between the states become stronger.

The $k=0$ cross-correlation is plotted in
Fig.~\ref{fig:CXL-hom-below}, and above the threshold there is a loss
of anticorrelation or there is even a small positive correlation.
This might be attributed to the macroscopic and independent
occurrences of the processes of step 2) and 3) in
Fig.~\ref{fig:basics}.

\begin{figure}[t]
  \begin{center}
 \includegraphics[width=8cm]{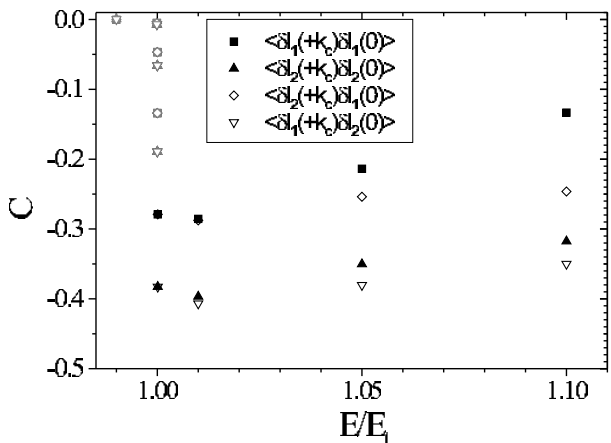}
 \caption{The correlations $C^n_{ij}(k=+k_c,k'=0)$ as function of the
   pump relative to the threshold. The gray symbols are the
   correlations below the threshold from Fig.~\ref{fig:CNLbelow}.}
 \label{fig:CNLabove}
\end{center}
\end{figure}

\begin{figure}[t]
  \begin{center}
 \includegraphics[width=8cm]{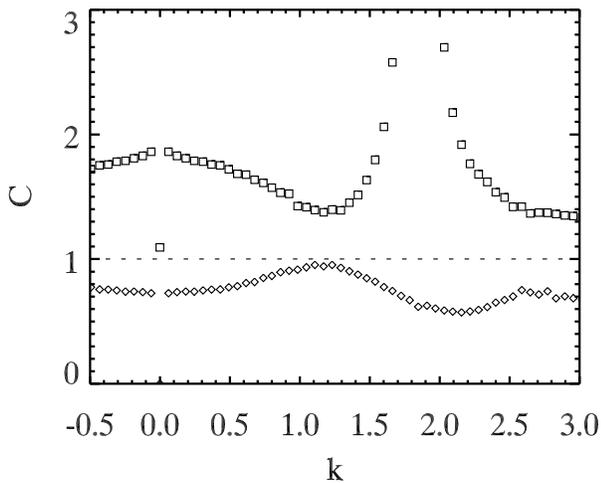}
 \caption{Photon number variances for $E/E_t=1.05$ showing
   $C_{11}^{(-)}(k,-k)$ (diamonds) and $C_{11}^{(+)}(k,-k)$
   (squares) from a numerical simulation.}
 \label{fig:CSN_FH_above}
\end{center}
\end{figure}

We saw in Sec.~\ref{sec:num-res} nonlinear correlations just below the
threshold, and in Fig.~\ref{fig:CNLabove} the peaks corresponding to
these correlations are plotted in order to follow the progress above
the threshold. The strongest anti-correlation is observed just above
the threshold, $E/E_t=1.0001$ and as the pump is increased the
correlations become weaker due to increasing competition of processes
involving higher harmonics. Moreover, the connection between the
self-correlations and cross-correlations seen below $E_t$ only remains
very close to the threshold, so as the pump is increased
$C^n_{11}(0,+k_c)\neq C^n_{12}(0,+k_c)$ and $C^n_{22}(+k_c,0)\neq
C^n_{12}(+k_c,0)$. This is related to the loss of perfect correlations
away from the threshold.

In Fig.~\ref{fig:CSN_FH_above} the photon number variances
$C_{11}^{(\pm)}(k,-k)$ above the threshold are presented. Comparing
these results with the corresponding ones below the threshold from
Fig.~\ref{fig:CSN_FH_below} we observe that they are very similar.
Generally, the correlation $C_{11}^{(-)}(k,-k)$ does not change much
with the pump level, and this fact has also been observed in the OPO
\cite{heidmann:1987}. The sum correlation $C_{11}^{(+)}(k,-k)$,
however, contains peaks that are very sensitive to the pump level,
both below and above the threshold. The behavior discussed here for
the FH is also valid for the SH and the cross-correlations.

\section{Conclusion and discussion}
\label{sec:conclusion}

We have used the master equation approach to describe the
spatiotemporal dynamics of the boson intra-cavity operators in
second-harmonic generation, and we included in the model quantum noise
as well as diffraction.  Our study is based on the Q representation to
describe the dynamics of the quantum fields in terms of a set of
nonlinear stochastic Langevin equations for equivalent c-number
fields. The choice of the Q representations gives some restraints on
the parameter space, but we have checked that similar results are
obtained using the approximated Wigner representation in other
parameter regions.

A simple scheme describing the microscopic photon interaction that
underlies the process of pattern formation has guided us in our
analytical and numerical studies of the spatial correlations. Equal
time correlations between intensity fluctuations were used to
investigate the strength of the correlations between different modes.
Also, possible nonclassical effects, such as twin beam correlations,
were considered by calculating the photon number variances of the
intensity sums and differences between spatial modes of the FH and SH
fields.

We have found that at the threshold for pattern formation the Fourier
modes with the critical wave number are perfectly correlated for the
FH field, the SH field and also between the FH and the SH field. As
the distance to the threshold is increased these correlations become
weaker, which was shown analytically to be due to the competition of
the eigenvalues of the linear system describing the system below the
threshold. At large wave numbers, only the correlation between
opposite points of the FH far field survives. This correlation is
always found to be stronger than the others, which is consistent with
the fact that the twin photon emission at the fundamental frequency is
the primary source for correlations in the system. For far field modes
around the critical wave number the self-correlations as well as the
cross-correlations between FH and SH photons are linked to the pattern
forming instability.

Very close to the threshold the linear analysis breaks down. The
numerical simulations below the threshold showed the existence of
nonlinear correlations which involve the $k=0$ mode and these are also
seen above the threshold. The other correlations described above are
also found above the threshold, but their strength decreases when moving
away from the threshold. This can be understood from the fact that
additional processes come into play, mainly consisting in population
exchanges between the macroscopic fields at the critical wave number
and its harmonics.

The intensity differences between opposite points of both the FH and
SH far fields, as well as the cross-correlation between the two have
been shown to exhibit nonclassical sub-shot noise behavior. These
properties for the intensity difference turn out not to be sensitive
to the process of pattern formation, since the corresponding
correlations depend very weakly on the distance to the threshold and
show no particular structure close to the critical wave number. The
emerging pattern is connected with increased fluctuations in the modes
with wave numbers around the critical wave number, leading to an
excess noise in the corresponding individual intensities. Therefore,
the sub-Poissonian statistics of the intensity differences reveal a
partial noise cancellation.  On the contrary, the sum of intensities
clearly exhibit peaks around the critical wave number, originating
from excess noise connected with the formation of a pattern.

In this work we considered equal time correlations calculated for the
intra-cavity fields. This approach turned out to be very useful to
understand the intra-cavity field dynamics. For the output fields we
expect that the nonclassical correlations of the intra-cavity fields
will remain below shot noise. The quantitative assessment of the
amount of noise reduction or excess noise with respect to the shot
noise level requires a specific additional calculation. For future
work it would also be interesting to calculate the output fluctuation
spectra at $0$ frequency for the difference and sum of intensities,
which reflect the full amount of quantum correlations induced by the
microscopic processes taking place inside the cavity, as for example
considered for a vectorial Kerr model in \cite{zambrini:2000}.

\section{Acknowledgments}

We acknowledge financial support from the European Commission projects
QSTRUCT (FMRX-CT96-0077), QUANTIM (IST-2000-26019) and PHASE and from
the Spanish MCyT project BFM2000-1108. We thank Steve Barnett and Pere
Colet for helpful discussions on this topic.

\bibliographystyle{c:/LocalTexMf/miktex/prsty}
\bibliography{c:/Projects/Bibtex/literature}

\begin{thebibliography}{10}
\bibitem{mathematica}
The steady states and eigenvalues are calculated through a semi-analytical
  treatment using Mathematica. This also holds for the rest of the analytical
  results.
\end{thebibliography}

\end{multicols}
\end{document}